# Analysis of Flow Field and Pneumatic Noise of The Flow around Series Cylinder

Yongxi Wu[1] *Guangnian Ji[2] Peizhi Zhang[3]
1 School of Mechanics and Engineering Science, Shanghai University, Shanghai 200072, China
2 Department of Mechanical, Materials and Manufacturing Engineering, The University of Nottingham Ningbo China, Ningbo 315100, China
3 Xingjian College, Tsinghua University, Beijing 100084, China
* Author to whom correspondence should be addressed: yongxiwu@connect.hku.hk

Abstract:Noise generation attributed to the flow around a cylinder and its control is of paramount importance in a multitude of engineering applications. In this study, we employ computational fluid dynamics (CFD) coupled with acoustic analogy, complemented by an analysis grounded in vortex dynamics theory, to explore the potential of wavy cylinder shapes in improving aerodynamic performance and mitigating aerodynamic noise. The series cylinder simplified from the aircraft landing gear and elliptical cylinder noise suppression mechanism is analyzed.Our research outcomes clearly indicate that elliptical ylinder designs have a notable impact on aerodynamic performance and noise reduction. Specifically, these designs effectively lower the average drag coefficient and efficiently suppress lift coefficient fluctuations, resulting in an overall reduction in noise production by the elliptical cylinder. To delve into the underlying mechanisms of noise suppression, our study meticulously examines the process of vorticity generation around the elliptical cylinder's surface. This examination yields profound improvements in the distribution of vorticity on the cylinder's surface, along with a remarkable weakening of the boundary vorticity flux and boundary enstrophy flux distribution. These changes result in a notable reduction in vorticity generation near the elliptical cylinder wall.These alterations directly lead to a substantial contraction in the distribution of vortex structures in the wake of the elliptical cylinder, especially affecting large-scale vortex structures.As a result,The elliptical cylinder noise suppression mechanism we propose can effectively reduce aerodynamic noise.

## 1. Introduction

In recent years, the stringent noise emission standards and heightened public concern regarding noise pollution have made noise reduction a critical consideration in the design of various modes of transportation, including aircraft, high-speed trains, and automobiles. Cylindrical components play a pivotal role in these industries, often being the source of significant aerodynamic noise pollution issues.For example, aircraft landing gear, primarily comprised of cylindrical components, has emerged as a prominent source of noise during the landing phase of large passenger aircraft. As a result, understanding the noise characteristics associated with flow around cylinders and developing noise reduction technologies has become of profound importance in addressing these challenges and enhancing various industrial applications.

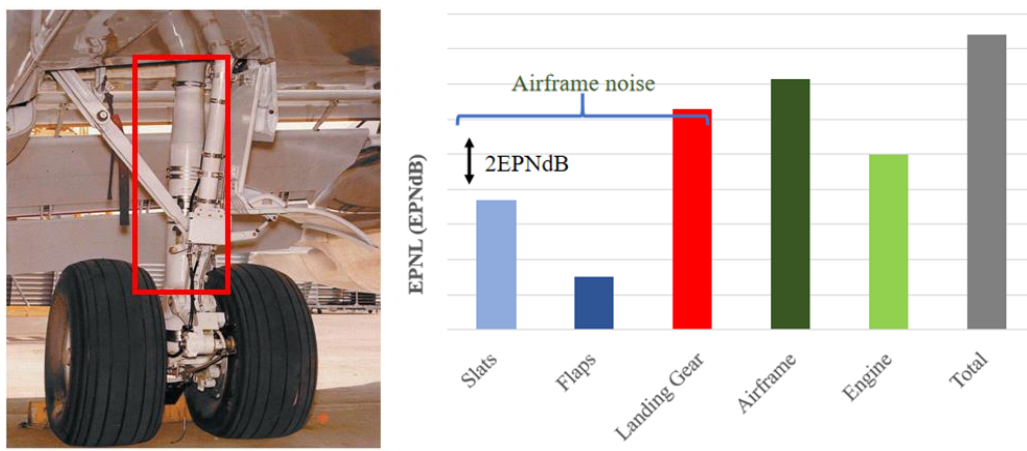

Fig1.1Series cylinders in aircraft landing gear(left)
1.2 The main noise source components of long range aircraft during landing

Cylindrical structures are associated with many practical applications, such as offshore risers, bridge piers, periscopes, chimneys, towers, masts, pillars, cables, antennas and wires. Understanding flow-related unsteady loads on such structures is critical to the design and control of hydraulics and aerodynamics [1], and there have been lots of previous work on this subject, repeated in articles published by Morkovin [2], Berger and Wille [3], Norberg [4], and Beaudan and Moin [8]. It is a benchmark problem in fluid dynamics for validating numerical and experimental techniques. Besides the basic understanding of the flow around cylinders, the researchers explored practical applications of this problem of the flow around cylinders, including offshore structures, heat exchangers, and fluid-solid coupling problems.

One of the main challenges in cylindrical winding research is to explore the formation and shedding mechanism of vortices in the wake region behind the cylinder. These vortices are caused by the flow interaction with the cylinders and are closely related to the Reynolds number. They play a critical role in determining the flow characteristics. Norberg [5] presented experimental results of measurements performed with cylinders of different diameters in cross fluid, and the Reynolds numbers range from 50 to $2 \times 10^5$.

In 1878, Strouhal investigated the wind squeak (Aeoliantones) produced by the flow around a cylinder and found that the frequency of the sound increased with the increase of wind speed but decreased with the increase of the cylinder diameter. In 1955, Curle [6] proved by theoretical analysis that the flow noise generated by flow around cylinder is a dipole sound source. The mechanism is mainly due to the unstable pressure generated by the interaction of the fluid with the surface of the cylinder. In 1956, Phillips [7] derived a formula for the sound intensity of noise when the flow is around a stationary uniform cylinder. In 1961, Gerrard [8] carried out an experimental study of lift and drag oscillations caused by the shedding vortex of a single cylindrical in a turbulent flow. In the same year, Roshko [9] investigated the flow noise around a single cylinder at high Reynolds numbers.

In addition to the study of aerodynamic noise mechanisms of flow around cylinder, for the purpose of reducing cylindrical noise, previous studies have proposed various active and passive control methods to improve the design and performance of these engineering applications including aircraft landing gears and heat exchangers. Most effective methods include injecting additional force and/or momentum into the low-energy boundary layer, such as synthetic jets, plasma actuators, air blowing, and suction. Typical passive methods include increasing surface roughness and adding wall perturbations, whose the main purpose is to retard vortex shedding as the boundary layer transitions from laminar to turbulent. Among passive control methods, porous materials have been heavily investigated due to their low cost and easy realization. Pikai Zhang et al. [10] applied highly porous materials to the separation zone in a subcritical fluid state (Re = $4.7 \times 10^4$). E. Latorre Iglesias et al. [11] evaluated the dependence of noise on yaw angle, flow velocity, cross-sectional shape, angle of attack, and radiation angle (directionality) using aeroacoustics wind tunnel tests for flow velocities ranging from 20 and 50 m/s of different cross-sections: circular, square, rectangular, and elliptical (Re = $1.6 \times 10^4$ to $1.2 \times 10^5$).

Researchers have been trying to understand the problem of disturbing flow of a double cylinder or even a row or rows of cylinders which the aircraft landing gear can simplified to.Most of the research on this topic has been based on experimental studies due to the complexity of the flow arising from the mutual coupling between the cylinders. Zhou and Yiu [12] investigated the wake structure and vortex dynamics for Re = $7 \times 10^3$ using hot-wire anemometry and identified two different flow structures in the reattached state. At $2 \leq L/D \leq 3$, shear layer reattachment occurs on the rear surface of the downstream cylinder, while at $3 \leq L/D \leq 5$, shear layer reattachment occurs on the front surface of the downstream cylinder. The interaction of a uniform flow with two closely placed objects produces strong non-constant aerodynamic forces, which cause severe non-linear vibrations and excessive noise. Flow interactions between airflow and tandem structures occur in many applications such as heat exchangers, twin cables in cable-stayed bridges, parallel bridges, and landing gear systems in airplanes. Therefore, the aerodynamics and aeroacoustics of tandem structures and have received extensive research attention. An important study of tandem cylindrical flows and noise was carried out at the NASA Langley Research Center in the Basic Aerodynamics Research Tunnel (BART) and the Quiet Flow Facility (QFF) [13][14][15]. In those experiments, the boundary layer flow was triggered to ensure that the laminar-to-turbulent transition occurs in the flow forward cylinder. In the tandem cylinders with a separation spacing of 3.7D, both cylinders showed detachment, producing strong interactions in the interstitial region. The conclusions demonstrated that this relatively simple geometry involves complex physics of noise generation [16]. Based on experiments conducted at NASA, Holami et al. [17] initially performed two-dimensional, fully turbulent simulations using the Unconstrained Reynolds Averaged Navier-Stokes (URANS) method to investigate the flow behavior around tandem cylinders and the mechanisms of noise generation. Furthermore, Bres et al. [18] utilized the lattice Boltzmann method (LBM) and the standard FW-H acoustic analogy to predict the flow and noise of tandem cylinders. Therefore, it is very important to use the CFD-based method coupled with acoustic analogy theory to predict the aerodynamic noise of single/tandem cylinders, to systematically carry out the generation of single/tandem aerodynamic noise sources and far-field noise analysis, to study the role of the Reynolds number on the flow and noise control, and to explore the new method of noise reduction for the tandem cylinders, to provide new ideas and new methods for the control of the noise. This paper contains five chapters, as follows: In Sec. II, the numerical methods are introduced. The theory of flow field computation and the theory of noise computation are described respectively. In Sec. III, the standard model is compared to verify the accuracy of the calculation, In Sec. IV, Sec. V, the results are compared to reveal the influence of high and low Reynolds number and series structure on the flow field and noise of cylindrical flow. Conclusions are made in Sec. VI

## 2. Numerical Methods

In this paper, compressible flow around single-cylinder and flow around tandem double-cylinder, and aerodynamic noise are numerically simulated by the IDDES simulation method using the CFD based on Finite Volume Method (FVM), and the flow-induced noise is derived by applying FW-H acoustic analogical analysis using SST $k$ - $\omega$ RANS model, with the temporal discretization format of second-order implicit, and spatial discretization using the Roe-FDS flux format, which is used for the flow-induced noise.

### 2.1　Aerodynamic Computation Methodology

Detached Eddy Simulation (DES) based on the S-A turbulence model was developed by Spalart et al. (1997) [19] to simulate flows with large-scale flow separation. It's a combination of RANS and LES formulations in DES, where RANS is used in the near-wall region and LES in the flow separation and wake regions, depending on the mesh resolution. To eliminate the possibility of model stress dissipation or mesh-induced separation in the original DES, Spalart et al. (2006) [20] proposed the Delayed DES (DDES). IDDES combines DDES and Wall Modeling LES (WMLES), proposed by Shur et al. (2008) [21]. In the IDDES formulation, the DDES length scale is used to eliminate the model stress loss of the original DES method, and WMLES is used to predict the average velocity of the boundary layer more accurately. In this study, the IDDES model based on the S-A model was used.

The key idea of IDDES based on the S-A model is to modify the length scale of the dissipation term in the S-A turbulence formulation (Spalart and Almaraz, 1994) [22], as given by Spalart et al. (1997). The S-A turbulence equation can be written as

$$\frac{d}{dt}\int_V \rho \tilde{v} dV + \int_S \rho \tilde{v} u ds = \frac{1}{\delta_{\tilde{v}}}\int_S (\mu + \rho\tilde{v})\nabla\tilde{v} ds + \int_V C_{b2}\rho(\nabla\tilde{v} \cdot \nabla\tilde{v})dV + \int_V (G_{\tilde{v}} - \gamma_{\tilde{v}} + S_{\tilde{v}})dV \quad (1)$$

where $V$ is the grid volume; $s$ is the oriented volume; $\rho$ is the fluid density; $\tilde{v}$ is the modified diffusion coefficient; t is the time; $\mu$ is the kinematic viscosity coefficient; $\delta_{\tilde{v}} = 2/3$; $C_{b2} = 0.622$; The first and second items on the right are diffusion items; $G_{\tilde{v}}, \gamma_{\tilde{v}}, S_{\tilde{v}}$ are the turbulence generation term, the dissipation term of eddy viscosity and the source term, respectively. The dissipation term $\gamma_{\tilde{v}}$ should satisfy

$$\gamma_{\tilde{v}} = \rho(C_{w1}f_w - \frac{C_{b1}}{k^2}f_{t2})(\frac{\tilde{v}}{\tilde{d}})^2 \quad (2)$$

where $C_{w1}$ is the equilibrium function and $f_{t2}$ is the model function; $f_w$ is the intermediate variable; $\tilde{d}$ is the length scale; $C_{b1}$ is 0.1335 and k is 0.41. In the original *DES* formulation, *d* is calculated by the following equation.

$$\tilde{d} = \min(d, C_{des}\Delta) \tag{3}$$

where $d$ is the distance to the nearest wall boundary, $\Delta=\max(\Delta x, \Delta y, \Delta z)$ is the maximum value of the grid spacing, $\Delta x$、$\Delta y$ and $\Delta z$ are local streamlines, wall normals, and transverse cell dimensions, respectively; $C_{des}$ 0.65. Due to the presence of some significant anisotropic meshing (i.e., $\Delta x \approx \Delta z \gg \Delta y$) in the meshing close to the wall, the DES length scale was used for the turbulence modeling throughout the boundary layer (i.e., $\tilde{d}=d$). In contrast, in the separation region or away from the wall, a subgrid-scale (SGS) model (i.e., $\tilde{d}=C_{des}\Delta$) was used instead of a turbulence model.

And when using a thin-walled parallel mesh or boundary layer thickening, it can be activated in the interior of the boundary layer $\tilde{d}=C_{des}\Delta$. This will reduce the eddy viscosity and lead to modeled stress losses in the boundary layer. On the other hand, if DES is used for wall model LES (WMLES), the simulation may result in a log-layer mismatch, which may lead to an underestimation of surface friction (Nikitin et al., 2000) [23]. To solve these problems, the IDDES model is proposed, which $\tilde{d}$ is defined as

$$\tilde{d} = \tilde{f}_d(1+f_e)d + (1-\tilde{f}_d)\Psi C_{des}\Delta_1 \tag{4}$$

where

$$\Delta_1 = \min(\max(0.15d, 0.15\Delta, \Delta_{\min}), \Delta) \tag{5}$$

$$\tilde{f}_d = \max((1-f_{dt}), f_B) \tag{6}$$

$$f_{dt} = 1 - \tanh[(C_{dt}r_{dt})^3] \tag{7}$$

$$r_{dt} = \frac{v}{k^2 d^2 \cdot \max\{[\Sigma_{i,j}(\partial u_i/\partial x_j)^2]^{1/2}, 10^{-10}\}} \tag{8}$$

where $\Delta_{min}$ is the minimum distance between the local grid center and the neighboring grid centers; $\psi$ is the low Reynolds number correction; $\upsilon$ is the kinematic viscosity coefficient; $f_B$ is an empirical co-mixture function; $C_{dt}=8$ and $C_{des}=0.65$; $f_e$ is the transformation function. If the simulation contains an inflow turbulence content of rdt≪1 and $\tilde{d}$ will be reduced to WMLES, which is defined as

$$l_{WMLES} = f_B(1+f_e)d + (1-f_B)\Psi C_{des}\Delta_1 \tag{9}$$

Otherwise, $\tilde{d}$ will be the DDES length scale lDDES which is defined as

$$l_{DDES} = d - f_d \max(0, d - \psi C_{des}\Delta_1) \tag{10}$$

where fd is the delay function. For more details, see Shur et al. (2008) [24].

## 2.2 Acoustic Computation Methodology

The Lighthill equation is obtained under the assumption of free space, so for places where solid boundaries do not play a role, such as the jet noise problem, Lighthill's basic theory is applicable. However, it has been proved that the effect of solid boundaries is of decisive significance for the sounding of stationary objects in turbulent flows and the sounding of moving objects. Therefore, it is of great practical significance to study the acoustic problem of fluid-solid boundary interaction. 1955, Curle extended Lighthill's theory to consider the effect of the stationary boundary. 1969, Ffowcs Williams and Hawkings extended Curle's result to consider the effect of moving solid boundaries on the acoustic field and obtained the famous FW-H equation. In the acoustic analog method, the obtained near-field flow is input as a sound source into the Ffowcs Williams and Hawking (FW-H) equation to predict the mid- and far-field noise. The FW-H equation can be obtained by controlling for continuity and the Navier-Stokes equation. In essence, it is a non-uniform fluctuation equation.

$$\frac{1}{c_\infty^2}(\frac{\partial^2 p'}{\partial t^2}) - \nabla^2 p' = \frac{\partial}{\partial t}\{[\rho_\infty v_n + \rho(u_n - v_n)]\delta(f)\} - \frac{\partial}{\partial x_i}\{[P_{ij}n_j + \rho u_i(u_n - v_n)]\delta(f)\} + \frac{\partial^2}{\partial x_i \partial x_j}\{T_{ij}H(f)\} \quad (11)$$

Where $p' = p - p_\infty$, $u_i$ and $v_i$ are the fluid and surface velocity components in the $x_i$ direction, and $u_n$ and $v_n$ are the fluid and surface velocity components perpendicular to the surface ($f = 0$). $\delta(f)$ denotes the Dirac function, and $H(f)$ is a mixture function. A mathematical surface (S) consists of a function $f$, $f<0$ denoting a region inside $S$, $f<0$ denoting the surface $S$, $f<0$ denoting an unbounded region outside $S$, and $n_i$ the unit normal vector pointing out of S, and $C_\infty$ is the free speed of sound. $T_{ij}$ is the Lighthill stress tensor. Equation 1A of Farassat[25][26] is the solution to the FW-H equation when the surface is moving at low Mach numbers, which ignores the quadrupole noise component. In this work, the volume component is neglected and only the area component is considered.

$$p'(x,t) = p'_T(x,t) + p'_T(x,t) \quad (12)$$

$$4\pi p_T'(x,t) = \int_{f=0}\left[\frac{\rho_\infty(\dot{U}_n+U_{\dot{n}})}{r(1-M_r)^2}\right]_{ret} dS + \int_{f=0}\left[\frac{\rho_\infty U_n\{rM_r+c_\infty(\dot{M}_r-M^2)\}}{r^2(1-M_r)^3}\right]_{ret} dS \quad (13)$$

$$4\pi p_L'(x,t) = \frac{1}{c_\infty}\int_{f=0}\left[\frac{\dot{L}_r}{r(1-M_r)^2}\right]_{ret} dS + \int_{f=0}\left[\frac{L_r - L_M}{r^2(1-M_r)^2}\right]_{ret} dS + \frac{1}{c_\infty}\int_{f=0}\left[\frac{L_r\{r\dot{M}_r+c_\infty(M_r-M^2)\}}{r^2(1-M_r)^3}\right]_{ret} dS \quad (14)$$

where the subscript *ret* denotes the value of the delayed moment; the dot superscript denotes the time derivative, the subscripts *T* and *L* denote the thickness noise and the load noise, respectively, and the variables with subscripts *r* or *n* denote the inner product of the vector and the unit vector of the radiation direction *r*, or the normal vector of the wall *n*, respectively, and $r_i$ is the vectorial diameter pointing from the acoustic source to the observation point, and $M_i$ is defined as $M_i = v_i/c_0$, and

$$l_r = l_i \cdot \frac{r_i}{r}, l_M = l_i \cdot M_i \tag{15}$$

$$M_r = M_i \cdot \frac{r_i}{r}, M_n = M_i \cdot n_i \tag{16}$$

$$\dot{M}_n = \dot{M}_i \cdot n_i, \dot{n}_M = \dot{n}_i \cdot M_i \tag{17}$$

$$\dot{M}_r = \dot{M}_i \cdot \frac{r_i}{r}, \dot{l}_r = \dot{l}_i \cdot \frac{r_i}{r} \tag{18}$$

Due to the low Mach number of the flow studied in this article, the quadrupole source contribution is not significant, so the quadrupole noise is ignored in this calculation, focusing on the monopole and bipole noise.

## 3. Computational Setup and Validation

### 3.1 Computational Setup

In this paper, we set up four sets of cases, case I is the standard case with Reynolds number *Re*=3900, which is the subcritical Reynolds number. Case II is the Trans-critical regime, compared with Case I to explore the effect of the Reynolds number on it. Case III is a tandem cylinder, and case IV is a tandem elliptic cylinder. Case III and case IV are compared to explore the effect of conformation.

For all the above cases, the Mach number is Ma = 0.2, e.g., 0.2 times the speed of sound. Cylindrical compressible wound flow and double-cylindrical tandem wound flow and aerodynamic noise are numerically simulated by the IDDES simulation method using the CFD tool based on Finite Volume Method (FVM), and the flow-induced noise is derived by applying FW-H acoustic analogical analysis using SST $k$-$\omega$ RANS model,

TABLE 1 List of all cases.

|  | Reynolds numbers(Re) | Flow regime | shape | Arrangement |
|---|---|---|---|---|
| case1 | $3.9\times10^3$ | Sub-critical regime | Cylinder | Single |
| case2 | $3.0\times10^6$ | Trans-critical regime | Cylinder | Single |
| case3 | $3.9\times10^3$ | 1.02 | Cylinder | Tandem |
| case4 | $3.9\times10^3$ | 0.99-1.02 | Ellipsoidal | Tandem |

The boundary conditions are set up as shown in the fig 3-1, with pressure far-field boundary conditions at the inlet, outlet, top and bottom, using periodic boundary conditions at the front and back, and non-slip wall boundary conditions on the cylindrical wall to simulate the winding of an infinitely long cylinder in the flow field. To fully capture the details of turbulent pulsation and improve the computational accuracy and convergence, a structured mesh with a grid number of about 290w is used, and the height of the first boundary layer is set to 0.0001m to ensure that the y$^+$ value of the wall is always less than 1. The 3D model of the computational domain is shown in the fig 3-2, the characteristic length, i.e., the diameter of the cylinder is D=0.05m, the flow direction is

12D in front of the cylinder, and 24D behind the cylinder, the width of the computational domain is 16D, the spreading size of the computational domain is set to be $Z=4D$, the blockage ratio is 6.25%, and the blockage ratio in the simulation is in the range of the values reported in the literature [27]. The effect of the spread size for $Z < 12D$ was mentioned in Weinmann [28] et al. It was confirmed that the effect of $Z$ on the numerical results is small. In numerical simulations, when the symmetry condition is applied in the spreading direction, it should be ensured that the size of the spreading domain is sufficient to capture all relevant flow structures [29].

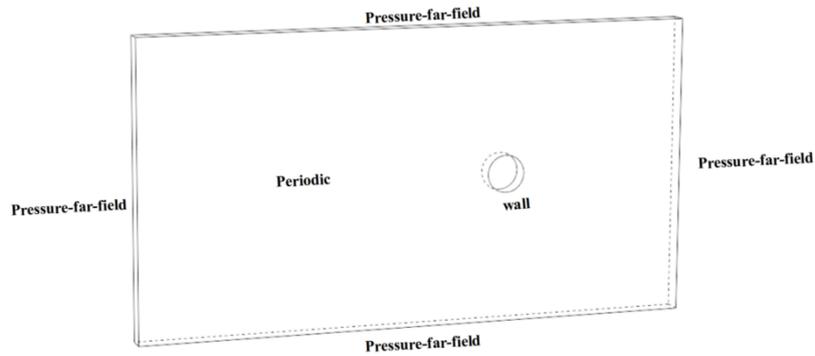

Fig 3-1 Schematics of boundary conditions

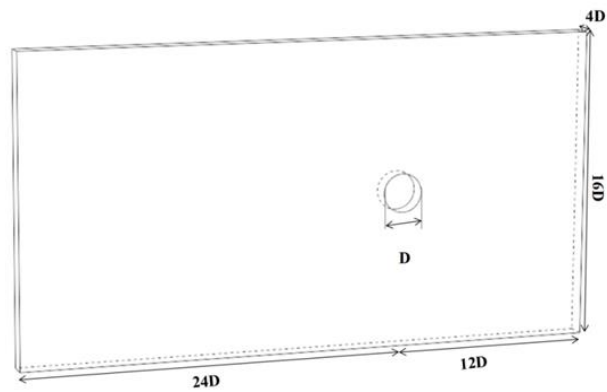

Fig 3-2 Schematics of the computational domain of single cylinder

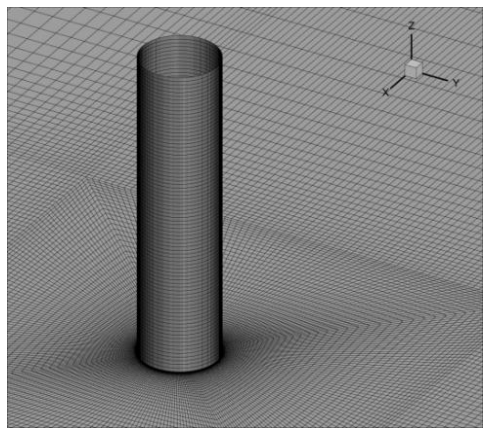

Fig 3-3 Grid distributions of the cylinder

For the tandem cylinders, the 3D modeling is as follows, the computational domain is the same as that of a single cylindrical winding, and the centroids of the two cylinders are separated by 3D. To find a noise reduction measure for the case of a double tandem cylinder arrangement, the upstream cylinder is changed to have an *AR*=1.6 ratio of the short to long axes, which is temporarily used for the study due to the time problem, and the *AR*=a/b is shown in Fig 3-5, and the other settings are the same as that of the tandem cylindrical winding.

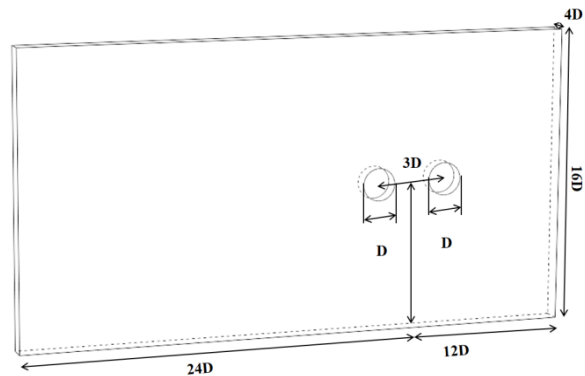

Fig 3-4 Schematics of the computational domain of tandem cylinders

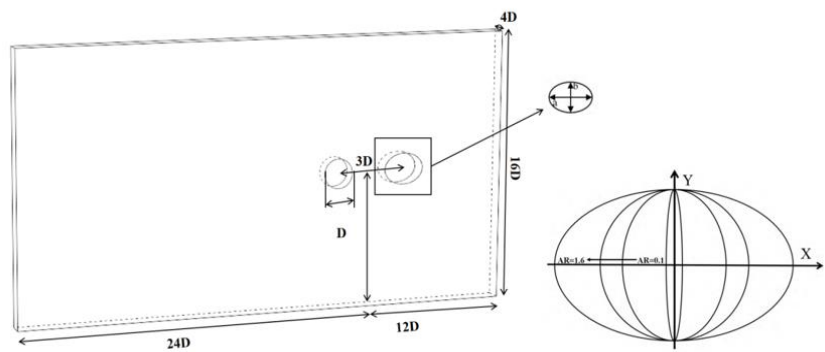

Fig 3-5 Schematics of the computational domain of tandem elliptical cylinders

The tandem cylindrical winding mesh setup is the same as the previous single-column winding setup, with the difference that the two cylindrical regions are encrypted individually so that the y+ values of both cylindrical walls are less than 1, and the total mesh amount is about 400w. The tandem elliptical mesh is also calculated using the same reliable mesh as the tandem cylindrical.

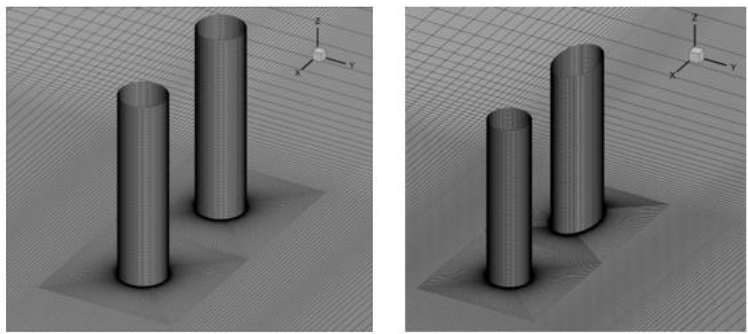

Fig 3-6 Grid distributions of tandem cylinders(lift) and tandem elliptical cylinders(right)

## 3.2 Numerical Validation

To verify the correctness of the numerical simulation results, the force coefficients and pressure distributions on a single cylinder were calculated with Re=3900. The numerical simulation results were also compared with the experimental data. As shown in the figure, the average drag coefficient, pulsation lift coefficient, and Strouhal number obtained from the simulation are compared with the experimental and simulation results, and they can correspond well, which indicates that the set of meshes can simulate the winding phenomenon related to vortex shedding very well.

TABLE 2 Comparison of main flow quantities.

|  | $Re$ | $\overline{C}_d$ | $C_{Lrms}$ | St |
|---|---|---|---|---|
| Present | $3.9 \times 10^3$ | 1.0473 | 0.0817 | 0.215 |
| Norberg(Exp) |  | 0.95-1.04 | 0.09 | 0.215 |
| Kahil |  | 1.02 | 0.137 | 0.207 |
| Frolich |  | 0.99-1.02 | 0.052-0.3 | 0.223 |
| Feng et al(LES) |  | 1.015 | 0.097 | 0.214 |
| Kravchenko&Moin(LES) |  | 1.04 | - | 0.210 |
| Present | $3.0 \times 10^6$ | 0.0727026 | 0.077 | 0.261 |
| Experiments |  | 0.36-0.78 | 0.05-0.13 | 0.17-0.29 |

To further validate the accuracy of the flow field simulations, a benchmark for numerical validation of the results with Re=3900 is compared with experiments and other numerical simulations. Average variables at baseline, such as time-averaged pressure coefficients $Cp$, and time-averaged velocities Ux=$U_\infty$ (dimensionless quantity), were compared with previous studies. The distribution of time-averaged pressure coefficients along the cylindrical surface is given in Fig 3-7, defined by $C_p = (p - p_\infty)/(0.5\rho_\infty U_\infty^2)$, where $p_\infty$ and $\rho_\infty$ are the pressure and density of the far-field incoming flow, respectively. The numerical results are in good agreement with the experimental data of Norberg [30] and the LES of Xu [31], Breuer [32] and Liyes [33].

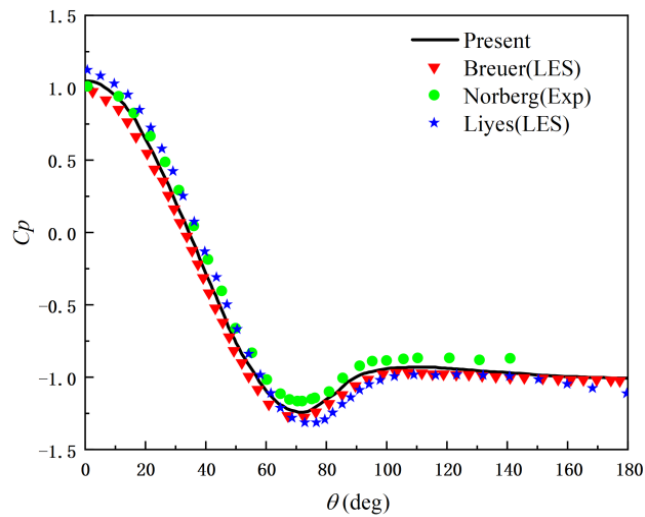

Fig 3-7 Time-average pressure coefficient on the wall surface of the cylinder

For the near-wall mesh density requirement, the mesh is encrypted to resolve the viscous sublayer in the near-wall region so that the y$^+$ value is always less than 1.0, as shown in Fig 3-8.

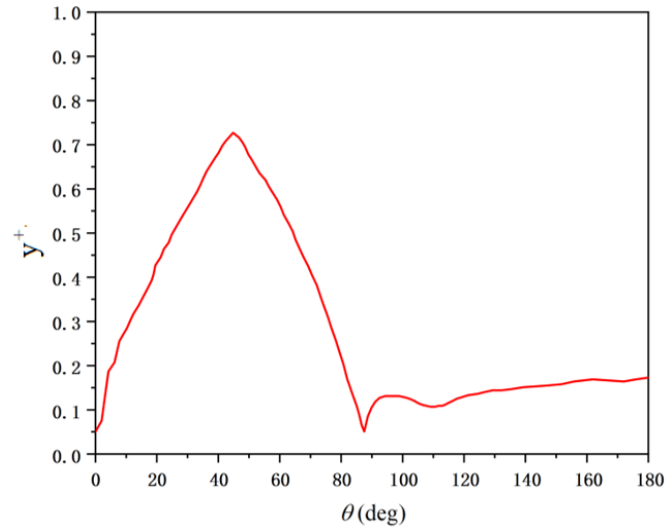

Fig 3-8 $y^+$ distribution for the cylinder.

As shown in Fig 3-9, the time-averaged flow velocity Ux=U∞ (dimensionless) along the centerline of the cylindrical wake is compared with the experimental and numerical results [34][35][36]. The results are in good agreement, but accurately simulating the flow development in the near wake region with a turbulence model is still a challenge due to the complexity of the flow in this region.

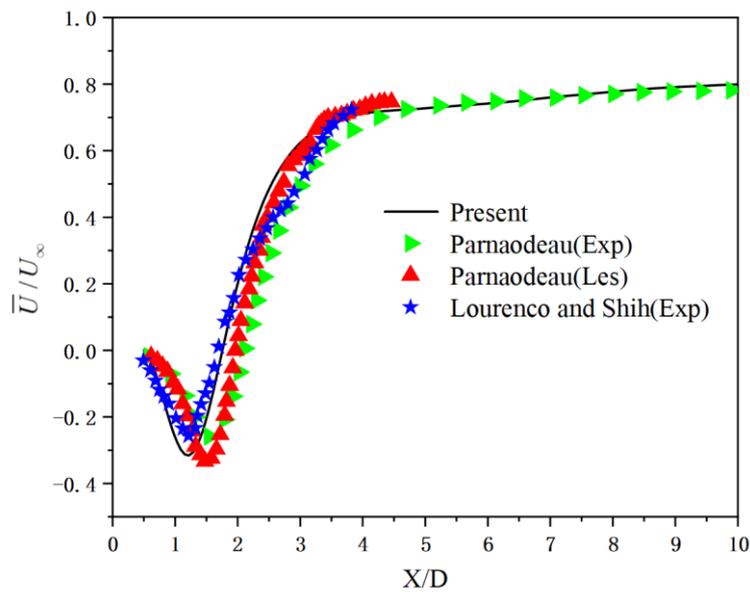

Fig 3-9 Time-averaged streamwise velocity along the centerline $Ux/U∞$

# 4. Results and Amalysis

## 4.1 Aerodynamic Analysis

As shown in Fig 4-1, the time-calendar lift coefficient and drag coefficient curves at Re=3900 and Re=3×10⁶, due to the larger flow rate, the lift-drag curve is no longer the most regular periodic curve, but the drag curve has maintained a level around the fluctuation of the lift coefficient shows a regular cyclic change. Comparing the force coefficients of subcritical Reynolds number and high Reynolds number, the lift coefficient is more fluctuant, and the overall lift is larger in the case of high Reynolds number due to the more complicated flow; at the same time, the drag coefficient is consistently larger than that of subcritical Reynolds number.

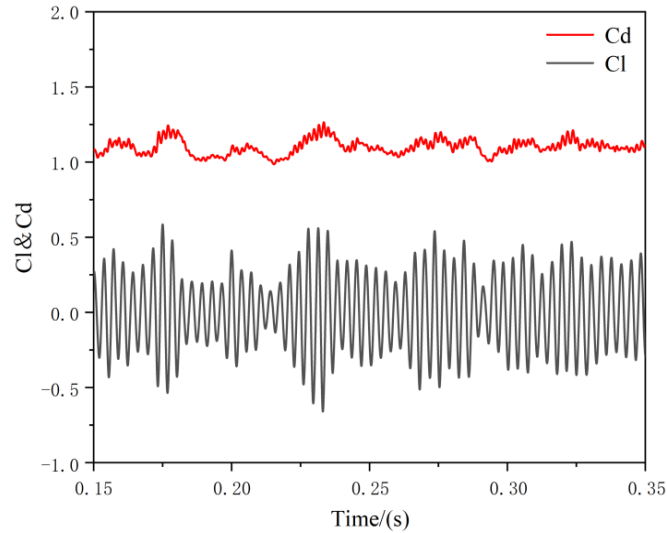

Fig 4-1 Drag coefficient and lift coefficient with Re=3900

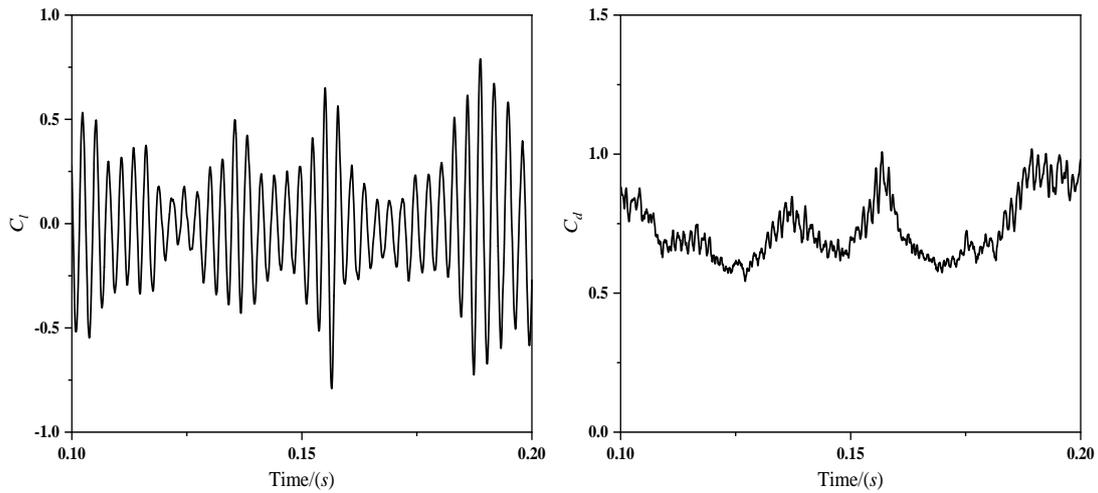

(a) lift coefficient with $Re=3\times10^6$          (b) drag coefficient with $Re=3\times10^6$

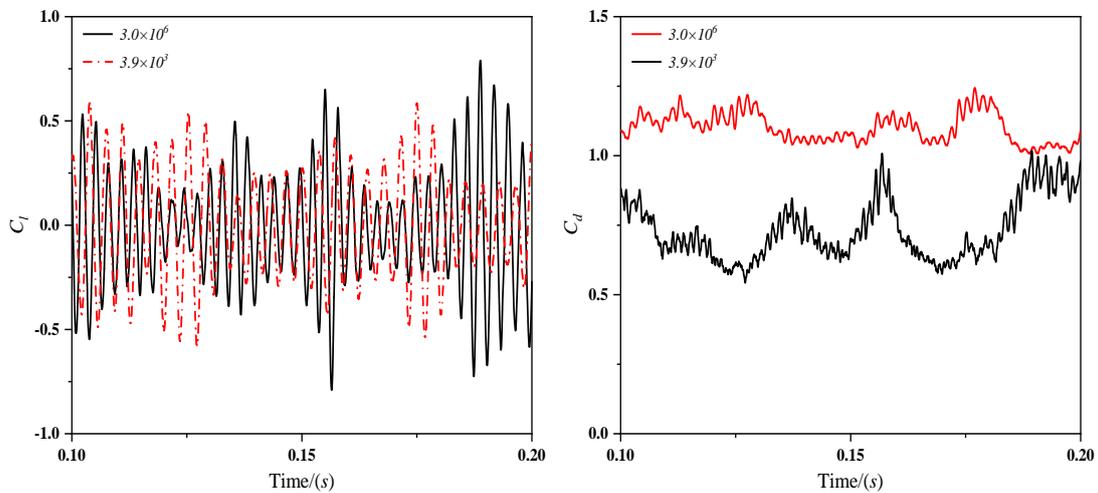

(c) Comparison of lift coefficient with $Re$=3900 and $Re$=3×10$^6$  (d) Comparison of drag coefficient with $Re$=3900 and $Re$=3×10$^6$

Fig 4-2 Comparison of drag coefficient and lift coefficient with $Re$=3900 and $Re$=3×10$^6$

In order to visualize and show more clearly the forces acting on the cylinders with time. In the following figures, we compare the phase space of the force components ($C_l$; $C_d$) at different Reynolds numbers ($Re$=3900 and $Re$=3×10$^6$). And it can be observed that the force components are in a reasonably small range for different cases. However, the fluctuations of the force components are suppressed more significantly at lower Reynolds numbers with a smaller range of fluctuations compared to the high Reynolds number conditions.

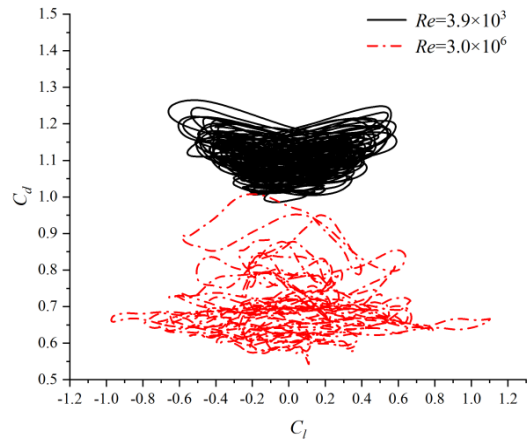

Fig Comparison of the phase-space plot of the force coefficient components for the cylinder with $Re$=3900 and $Re$=3×10$^6$

The force on the cylinder is determined by the pressure and viscous shear stress. Therefore, in order to further analyze the force distribution of the disturbed flow on the cylinder, the following figures show the shear stress clouds of the cylinder at different Reynolds numbers ($Re$=3900 (left) and $Re$=3×10$^6$ (right)), respectively. By observing the following figure (left), it can be clearly known that the shear stresses on the middle and two side parts of the cylinder are the largest. But when $Re$=3×10$^6$, the shear stress is greatly reduced and the distribution is chaotic. According to the previous analysis of Fig4-2(d), when $Re$=3×10$^6$, the fluctuation of the cylindrical drag coefficient is almost kept at a level. This also means that in this case, the fluid separates quickly after contacting the cylinder and forms wake vortices from the disturbed flow over the upper and lower surfaces of the cylinder. So it did not cause large viscous shear stress on the surface of the cylinder.

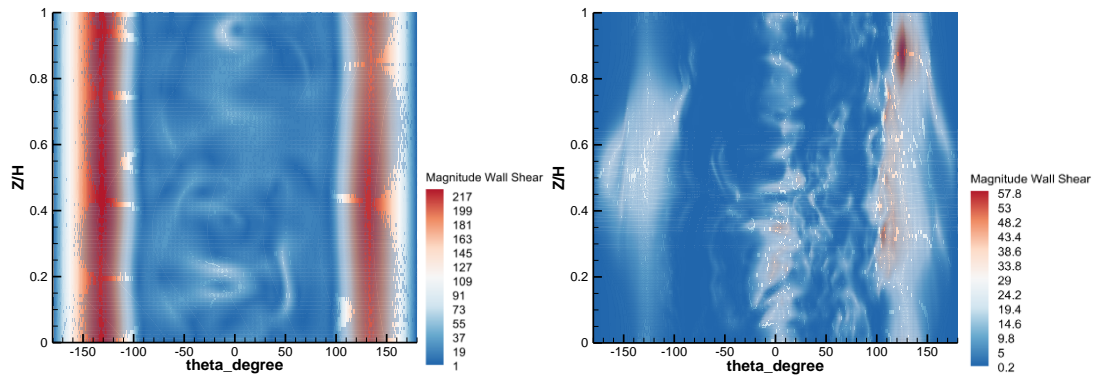

Fig The wall shear stress of single cylinder with $Re$=3900 (left) and $Re$=3×10$^6$(right)

For the Strouhal number St, which expresses the frequency of vortex shedding, the larger the Reynolds number the faster the vortex shedding and the larger the St. The larger the Reynolds number, the faster the vortex shedding.

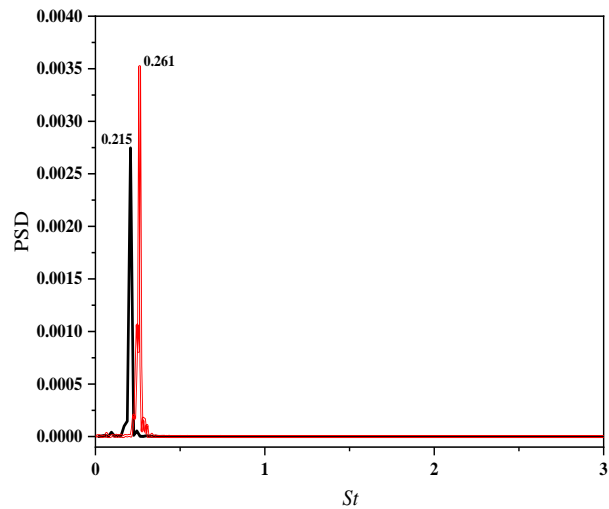

Fig 4-3 Comparison of St number with Re=3900 and Re=3×10⁶

## 4.2 Flow Field Analysis

As shown in Fig 3-4, the time-averaged streamwise velocities Ux=U$_\infty$ (dimensionless quantity) at $Re$=3900 and at $Re$=3×10$^6$ along the centerline of the cylindrical wake were compared. The trend of the wake flow distribution is the same for both Reynolds numbers. And in the wake region near the cylinder, the time-averaged flow velocity decreases with flow development due to vortex shedding on the cylinder surface. After reaching a minimum in the turbulent core, it recovers with the development of downstream flow. With further development of the wake, the time-averaged velocity will recover to the same level as the upstream average velocity. There are some changes in the velocity distribution in the wake region at different Reynolds numbers. The location of the minimum velocity at low Reynolds number moves further upstream, indicating that the core region of the turbulence is located further downstream of the wake zone, which can be further verified in the subsequent surface streamline chart.

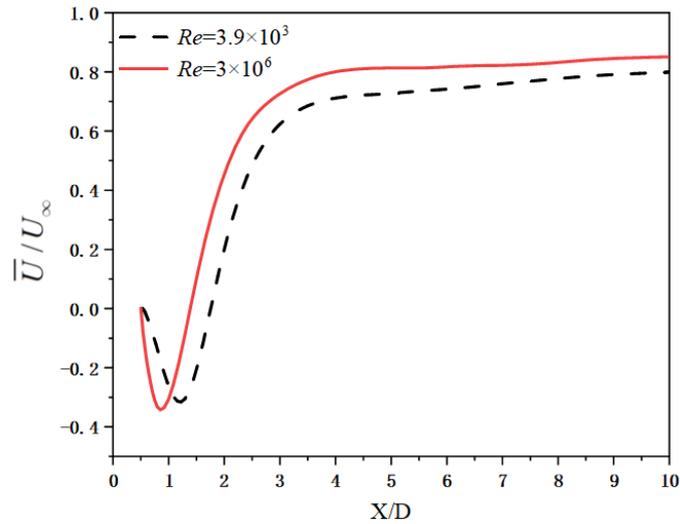

Fig 4-4 Comparison of time-averaged streamwise velocity Ux=U∞ (dimensionless quantity)

along the centerline of the cylindrical wake at Re=3900 and Re=3×10⁶

Observing the time-averaged pressure stress coefficient program of the cylindrical surfaces and their develop diagram at two different Reynolds numbers with a huge difference, a large pressure gradient occurs on both surfaces and this pressure gradient increases with the larger Reynolds number; a larger negative pressure coefficient occurs on the wall surface in the case of the high Reynolds number.

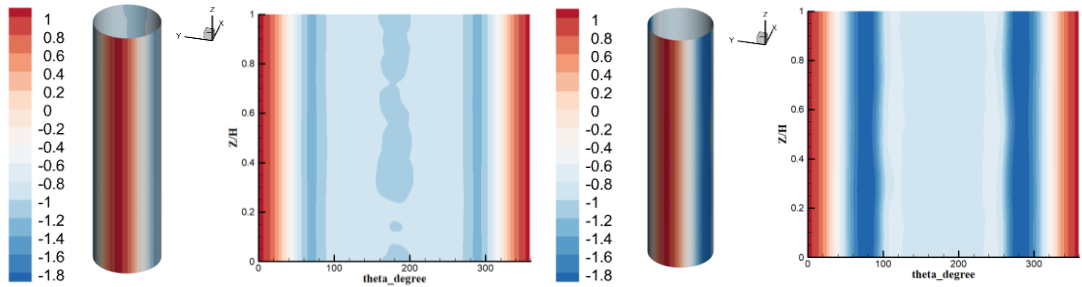

Fig 4-5 The time-averaged pressure stress coefficient program of the cylindrical surfaces

and develop diagram at Re=3900(lift) and Re=3×10⁶(right)

As shown in Fig 3-6, the higher the Reynolds number the greater the pressure change and the more chaotic the distribution of the surface pressure rate of change, compared to the low Reynolds number case where the pressure rate of change is greatly reduced.

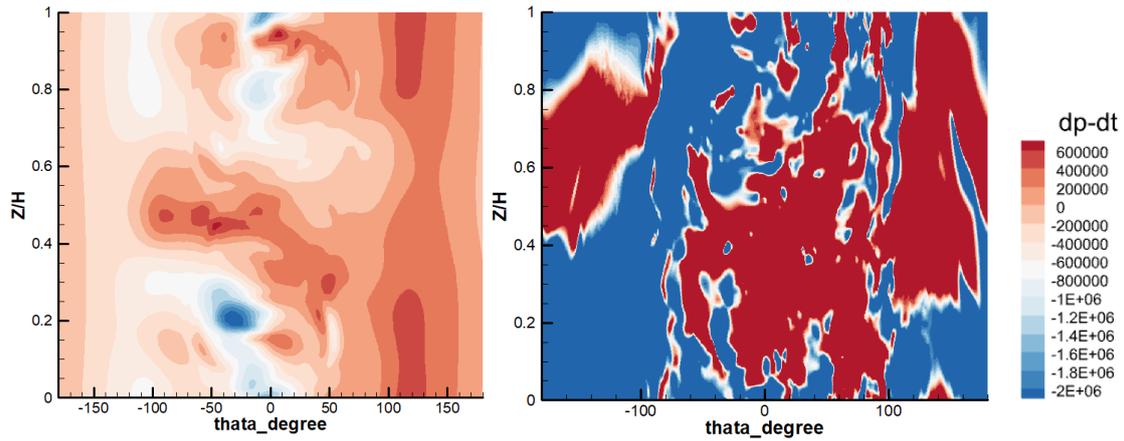

Fig 4-6 Develop diagram of the surface at Re=3900(lift) and Re=3×10⁶(right)

Figure 4-7 shows end view vortical structures by the Q-criterion(Q = 1000), rendering at X-direction. In both cases, there is a clear periodic Kamen vortex structure behind the cylinder. As the Reynolds number increases, the separation position of the cylindrical surface moves downward. By comparison, it can be observed that the shear layer on the low Reynolds number surface is more stable than the high Reynolds number in both sets of Reynolds number cases, and the high Reynolds number is unstable further downstream, translating into small-scale vortex structures. The distribution of vortex structures in the flow field of the low Reynolds number cylinder is more concentrated, and the large-scale flow vortices, hairpin vortices and other structures are also significantly reduced, and the vortex strength of the wake is significantly weakened, which makes the influence of the wake on the fluctuation of the column surface weaker.

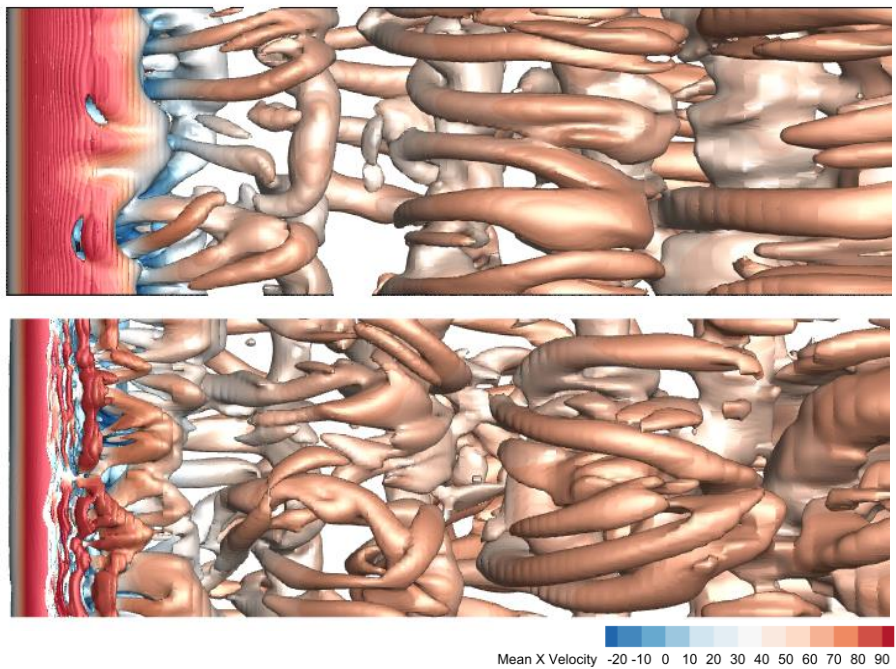

Fig4-7 End view of vortical structures by the Q-criterion(Q = 1000) at Re=3900 and Re=3×10⁶

Fig 4-8 shows the vorticity by the Q-criterion(Q = 1000), rendering at X-direction. The distribution of vortex structures in the low Reynolds number cylindrical flow field is more concentrated, and the large-scale flow vortices, hairpin vortices and other structures are also significantly reduced, and the vortex strength of the wake is significantly weakened, which makes the influence of the wake

on the fluctuation of the column surface weakened.

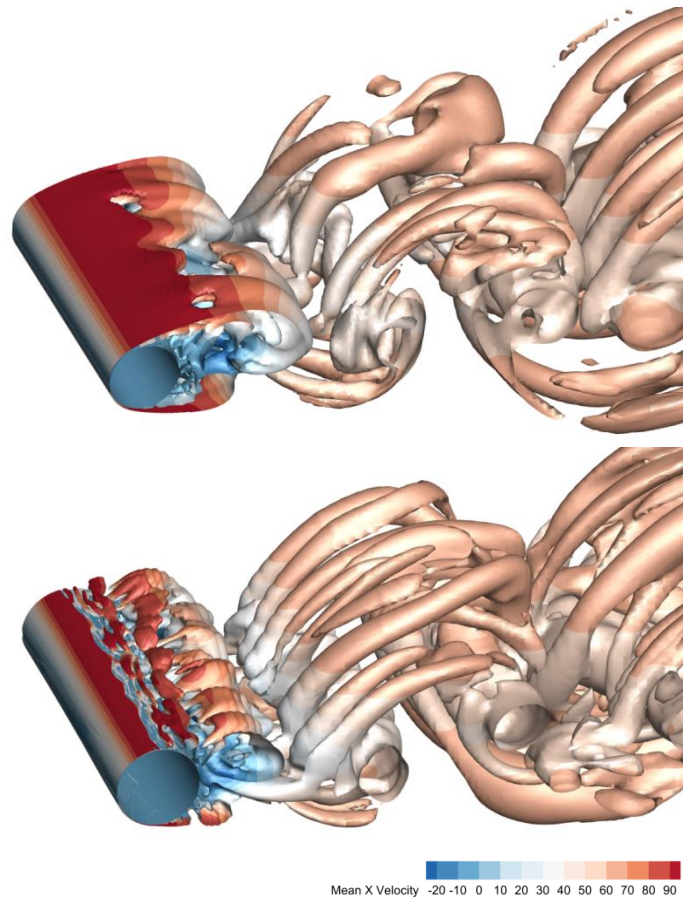

Fig 4-8 Vortical structures by the Q-criterion(Q = 1000) at Re=3900 and Re=3×10⁶

As shown in Fig 4-9, a comparison of vortex structures for Re=3900 and Re=3 × 10⁶ shows that for the high Reynolds number case, there is more vortex structure, shorter elongation distances from the wall, and a greater concentration of vortices near the wall.

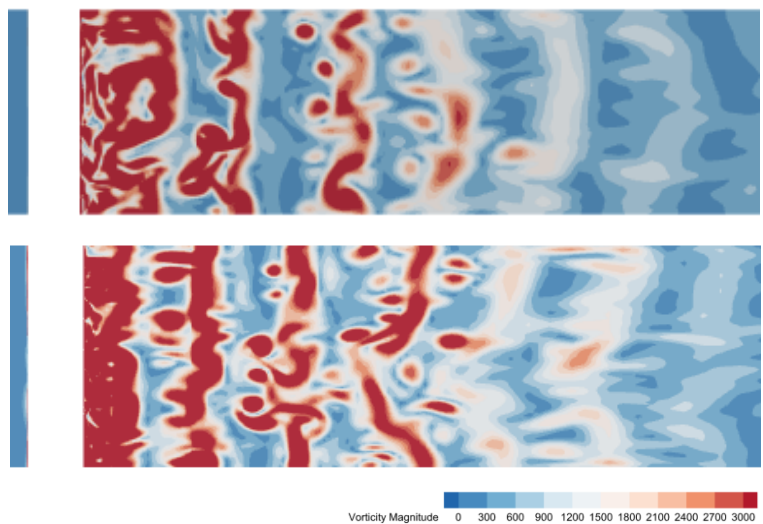

Fig 4-9 End view of vortical structures by the Q-criterion(Q = 1000) at Re=3900 and Re=3×10⁶

Figures 4-10 and 4-11 show the time-averaged streamwise velocity distribution contours compared to the streamlines. At low Reynolds numbers the shear layer of the cylinder is more stable, and its instability occurs farther downstream, which leads to the extension and enlargement of the wake

region and downstream motion of the turbulent core region. It can also be observed that at high Reynolds number, the cylindrical surface shows larger velocities and velocity gradients, and the velocity streamlines are narrower and more elongated in area.

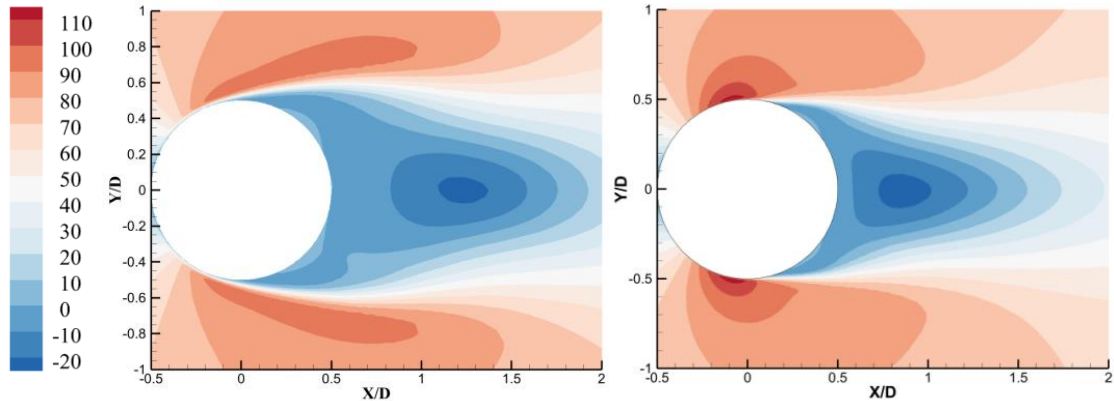

Fig 4-10 Time-averaged velocity program (dimensionless quantity) in the x-direction at Re=3900 (left) and Re=3×10$^6$ (right)

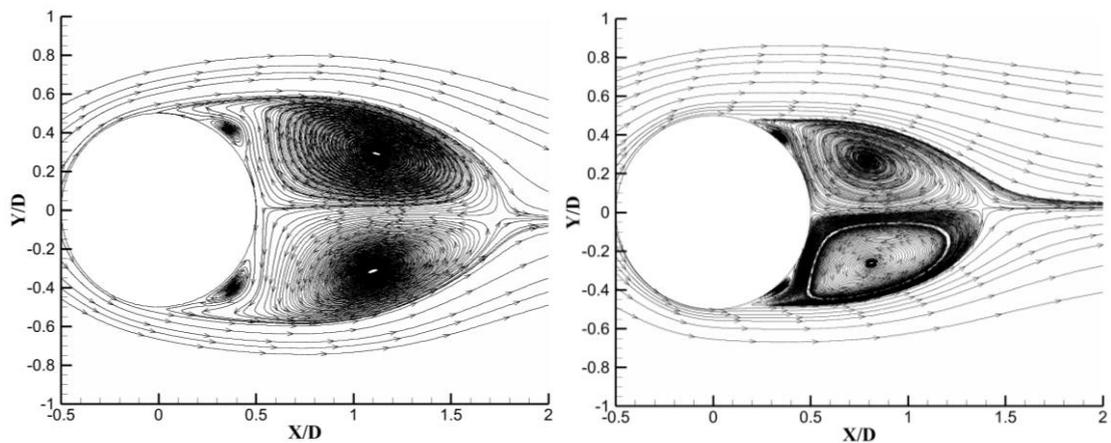

Fig 4-11 Streamline charts(dimensionless quantity) at Re=3900 (left) and Re=3×10$^6$ (right)

To further understand the differences in flow physics at Re=3900 and Re=3×10$^6$, we investigated wake turbulence and turbulence intensity. The wake turbulence intensity can be quantified by the turbulent kinetic energy in Fig 4-12. As shown in the figure, a higher level of turbulent kinetic energy occurs in the wake region of the upstream cylinder at Re=3×10$^6$. In addition, at Re = 3900, the turbulent viscosity level in the vortex formation region is higher than the shear layer on the upstream cylinder. In contrast, at Re=3×10$^6$, the highest turbulent kinetic energy is mainly concentrated in the shear layer on the upstream cylinder rather than in the vortex formation region.

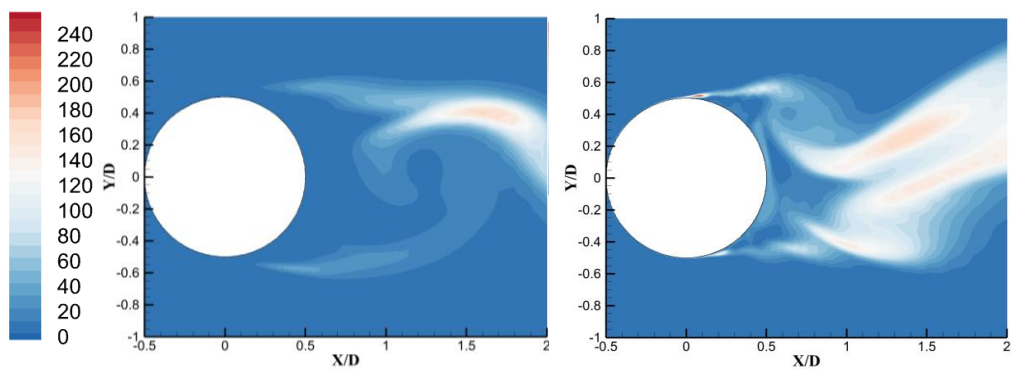

Fig 4-12 Turbulent kinetic energy(dimensionless quantity) at Re=3900 (left) and Re=3×10$^6$ (right)

Fig 4-13 demonstrates the comparison of the surface-limiting streamlines at Re=3900 and Re=3×10$^6$. The results show that the surface-limiting flow lines of cylinders at low Reynolds number are more regular than the spreading distribution of circular cylinder, and the distribution of separation lines at high Reynolds number is more chaotic in terms of quadrilateral-like and irregular distribution, and the separation point of the flow at high Reynolds number can be seen to be more hysteretic from the end view.

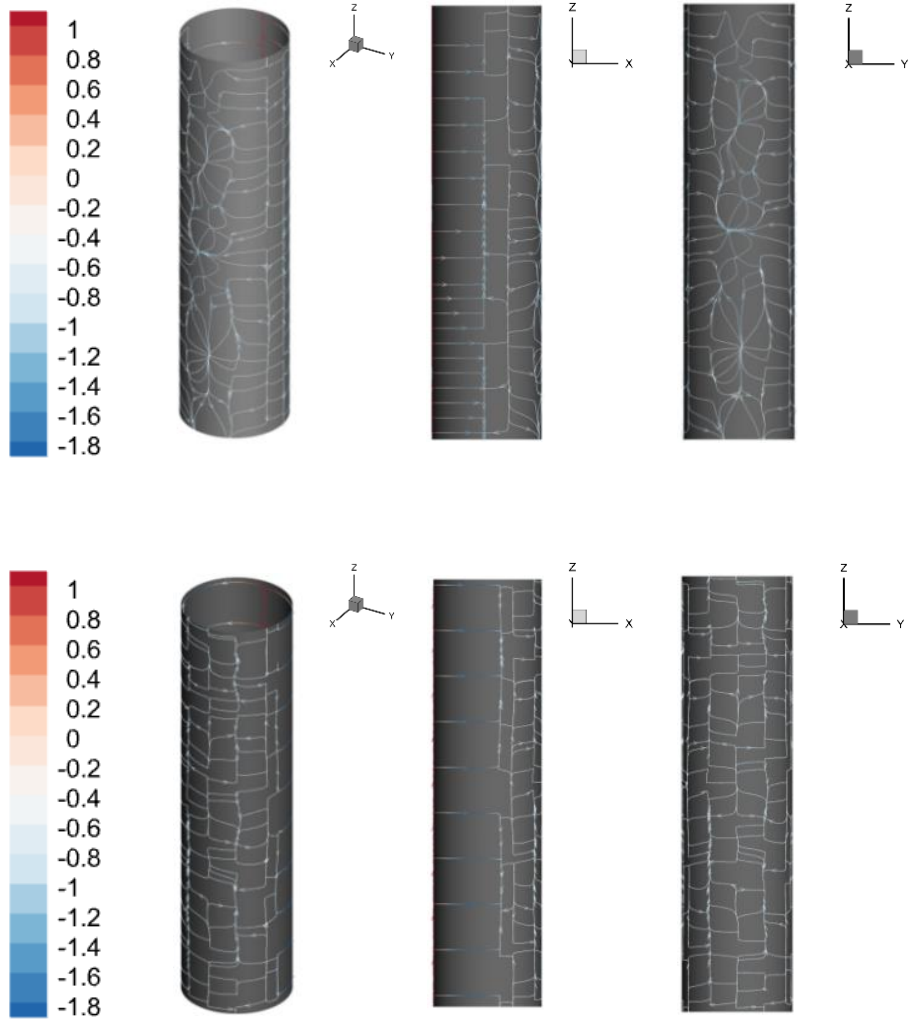

Fig 4-13 The surface limiting streamline distribution(rendering at X-direction) at Re=3900 (former) and Re=3×10$^6$ (later)

## 4.3 Analysis of Far-field Radiated Noise

Due to the low free-speed Mach number (Ma = 0.2), the contribution of the quadrupole source is not very significant; therefore, most of the sound is produced near the wall (dipole and monopole sources). Therefore, all noise sources generated on the surface of the cylindrical wall are considered when calculating the acoustics in this paper. The far-field noise spectrum was obtained by FW-H integration and Fast Fourier Transform (FFT), as shown in Fig 4-14, using the observer position at 40D from the cylindrical axis, located at 90° from the stationary point of the cylinder as a comparison with the experiments to validate the accuracy of the acoustic field calculations. The overall sampled data was subdivided into segments with 18614 sampling points per segment. A hanning window (50% overlap) was used and the results were averaged over all windows. The frequencies of the main frequency noise are almost the same, and the other frequencies are also in good agreement, the main frequency obtained from the simulation is about 84.19 Hz, and the experimental main frequency is 84.68 Hz, with an error of 0.6%. After that, 24 noise receivers are uniformly distributed along the perimeter of the cylinder, with the first number of receivers at 40D to the rear of the flow around cylinder, which is rotated counterclockwise in turn, and finally, the OASPL for each observer is obtained by calculation.

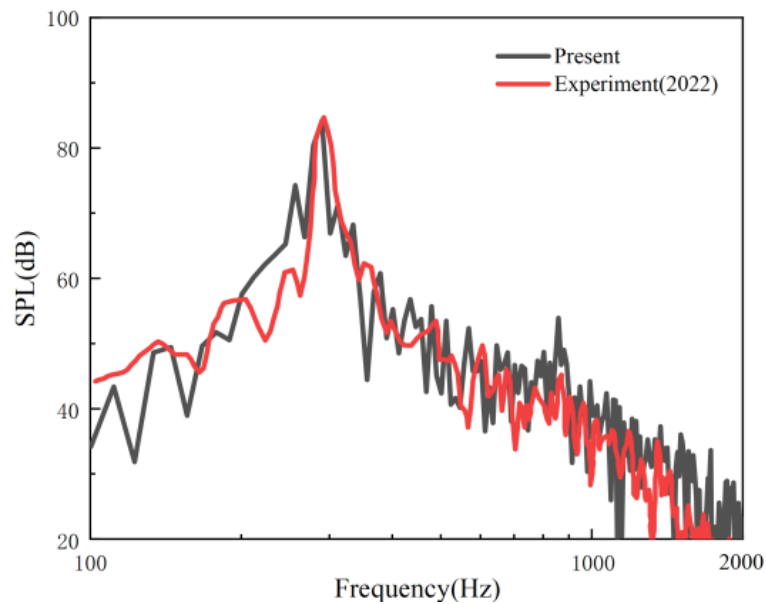

Fig 4-14 Acoustic pressure spectra

Figs 4-15, 4-16, and 4-17 contain the sound pressure level-frequency curves at the complete 24 observation points, and it can be seen that except for the observation point at 0° and 180° where the main frequency exceeds 500 Hz, the main frequencies of the other observation points are all below 500 Hz, and the low frequencies are dominant, and the higher the frequency, the lower the sound pressure level. In addition, in some observation points, there is a sub-major frequency other than the main frequency, and the sound pressure level difference with the main frequency is not big. As for the noise distribution at high Reynolds number, it can be seen that the distribution of low frequency and high frequency is more balanced, but the low frequency still occupies a part of the dominant position, also in the observation points at 0° and 180°, the main frequency exceeds 500Hz, and the main frequency of the other observation points are all below 500Hz, and it is worth noting that, at the Reynolds number of such a big difference, the difference in the SPL of the main frequency is not significant.

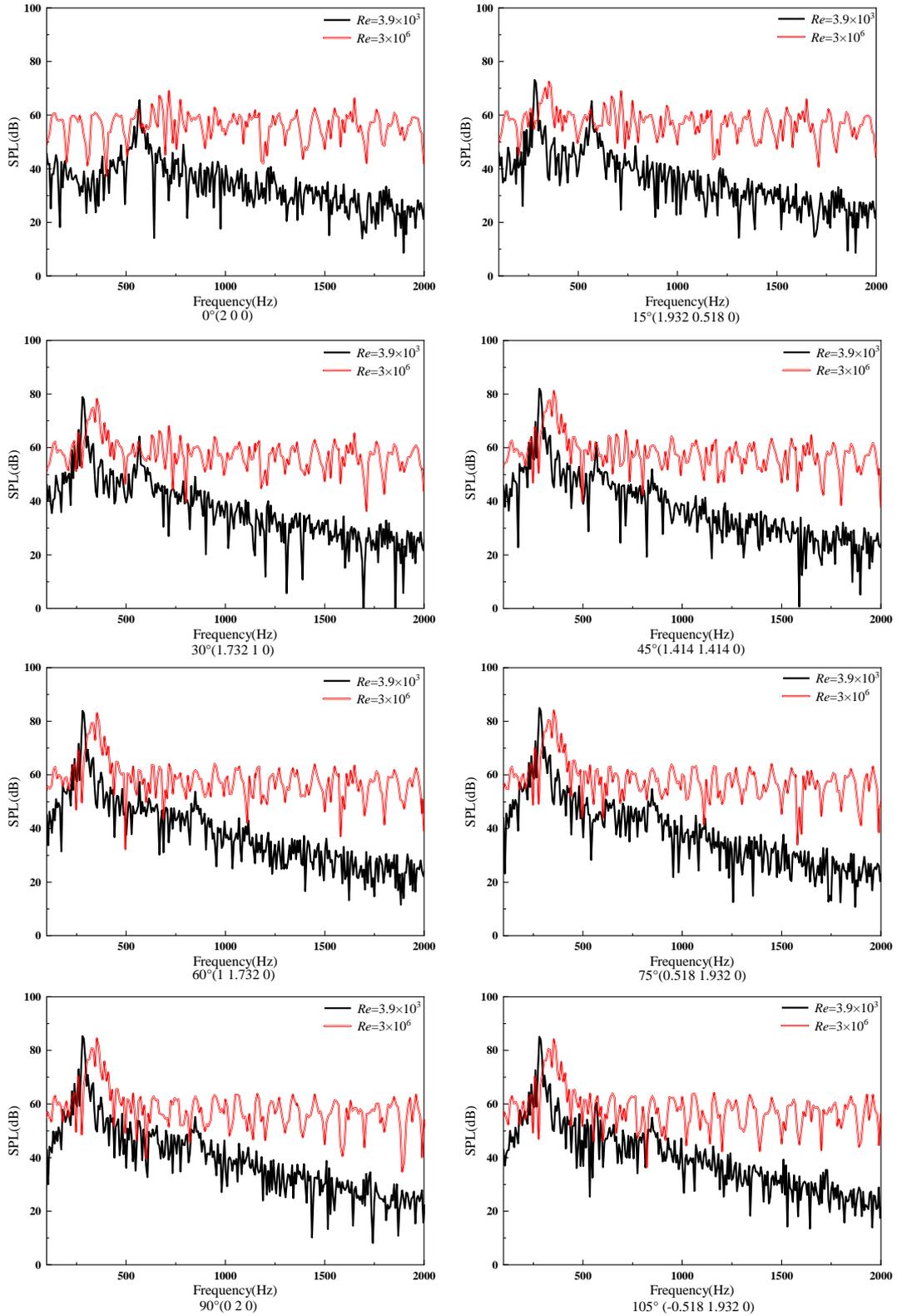

Fig 4-15 Comparisons of acoustic pressure spectra with the frequency of the observation points 1-8 of the single cylinder winding at Re=3900 and Re=3×10$^6$

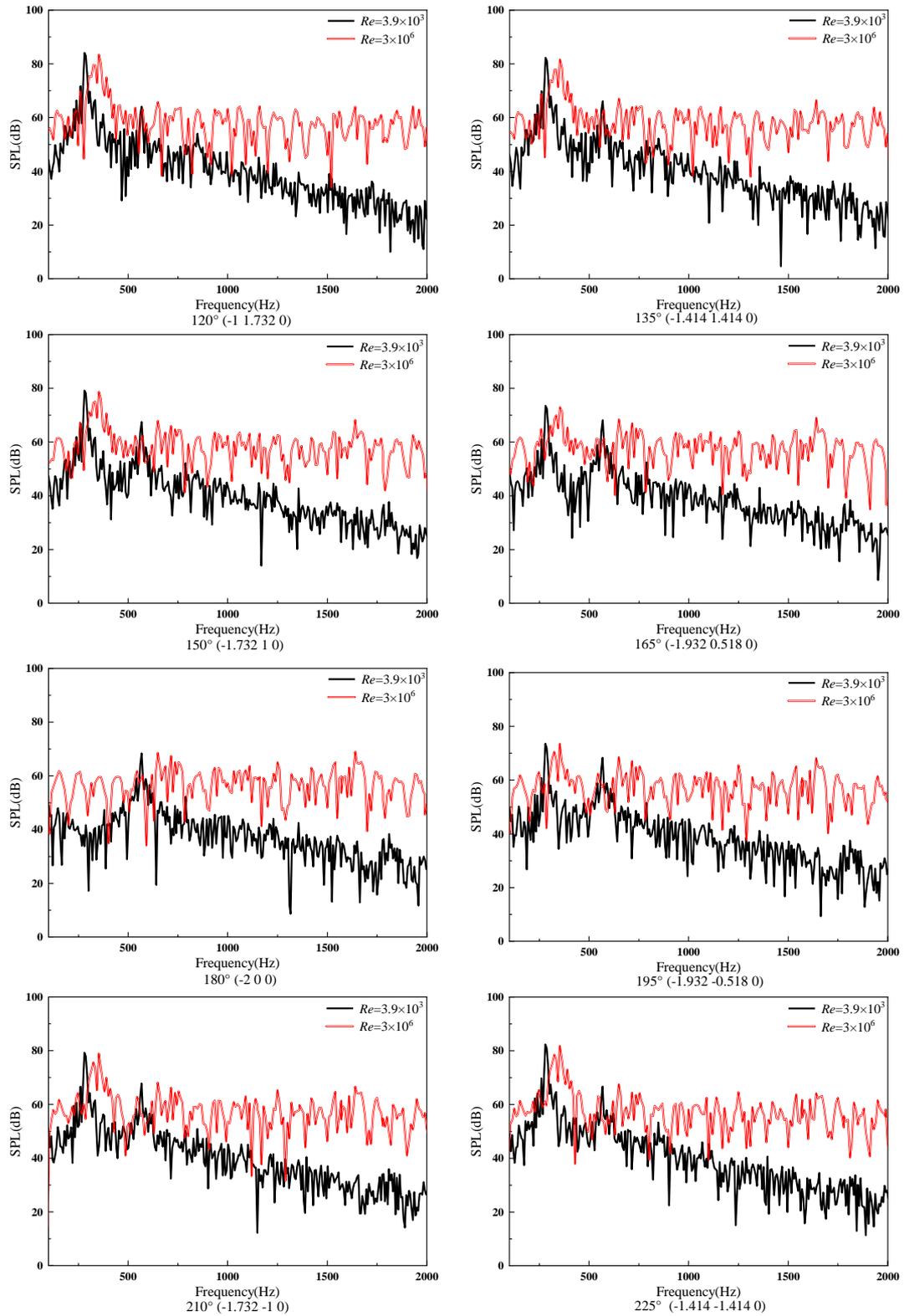

Fig 4-16 Comparisons of acoustic pressure spectra with the frequency of the observation points 9-16 of the single cylinder winding at Re=3900 and Re=3×10⁶

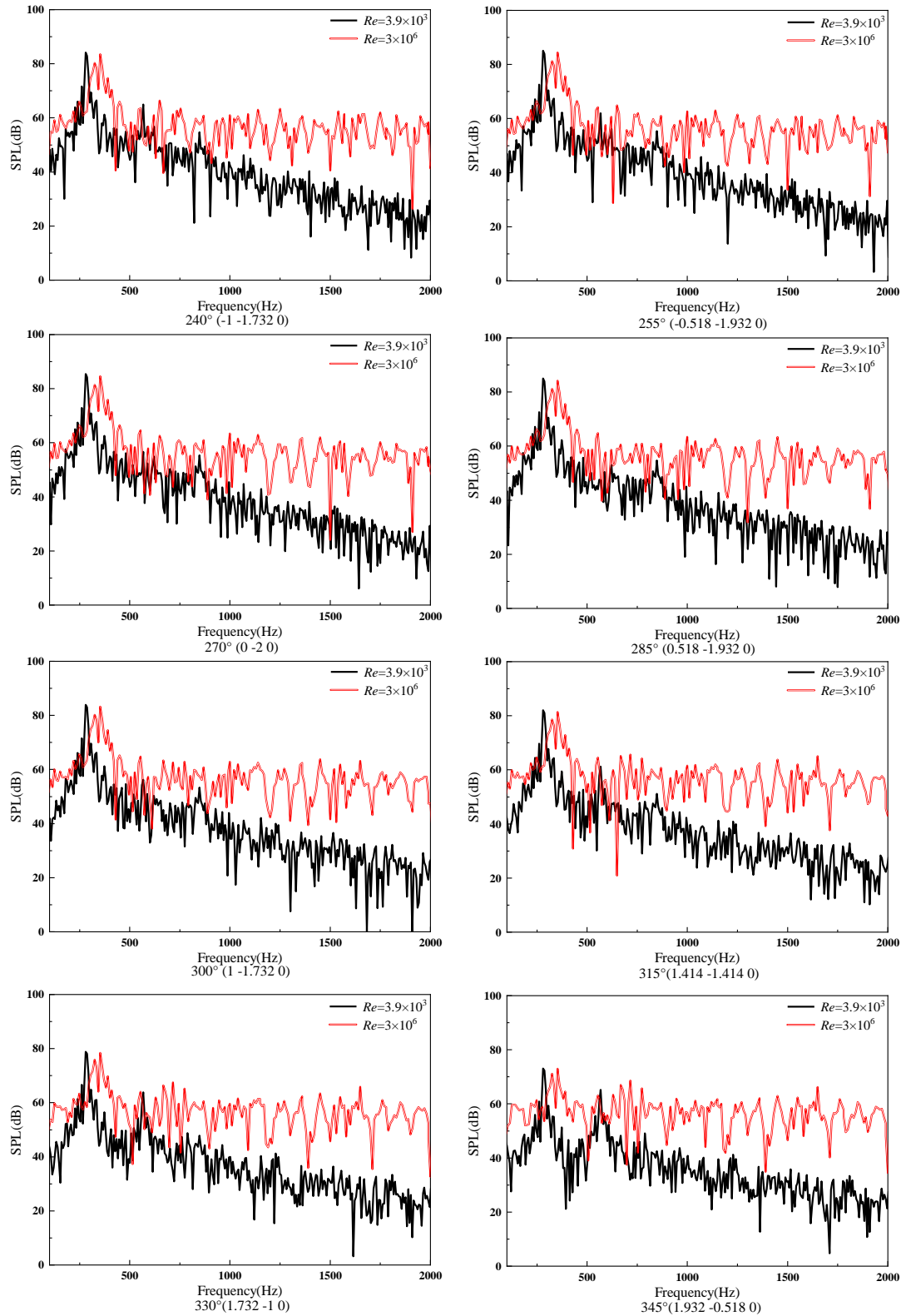

Fig 4-17 Comparisons of acoustic pressure spectra with the frequency of the observation points 17-24 of the single cylinder winding at Re=3900 and Re=3×10⁶

Based on the uniform distribution of 24 observation points along the perimeter of the cylinder, the total sound pressure level OASPL for each observer is obtained by calculation and the sound directivity distribution is realized as in Figure 3-26. In the shape of the noise directivity is consistent for both Reynolds numbers and both have dipole modes, the noise radiation increases in all directions for high Reynolds numbers, with the most pronounced increase in noise at 0° and 180°, and the least increase in noise perpendicular to the direction of flow.

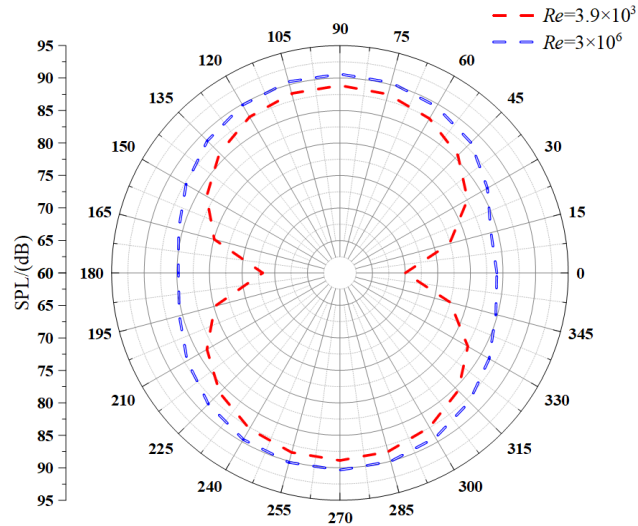

Fig 4-18 Comparison of the directivity of noise at Re=3900 and Re=3×10⁶

## 5.1 Aerodynamic Analysis

The lift resistance coefficients of the front and back cylinders are shown in Fig 5-1, and the lift resistance coefficients of the front cylinder do not have large fluctuations and are not periodic, which are related to the vortices on its surface not being completely dislodged, with an average drag coefficient of 0.8. The lift resistance coefficients of the back cylinder, on the other hand, have large fluctuations and are periodic, with the lift coefficients peaked at about 0.5, and the average drag coefficients are -0.2. This is because Cl fluctuations in the front cylinder are caused by the motion of the shear layer, whereas Cl fluctuations in the back cylinder are caused by vortex shedding, and for L/D = 3.0, a bistable flow pattern occurs and affects vortex shedding behind the back cylinder.

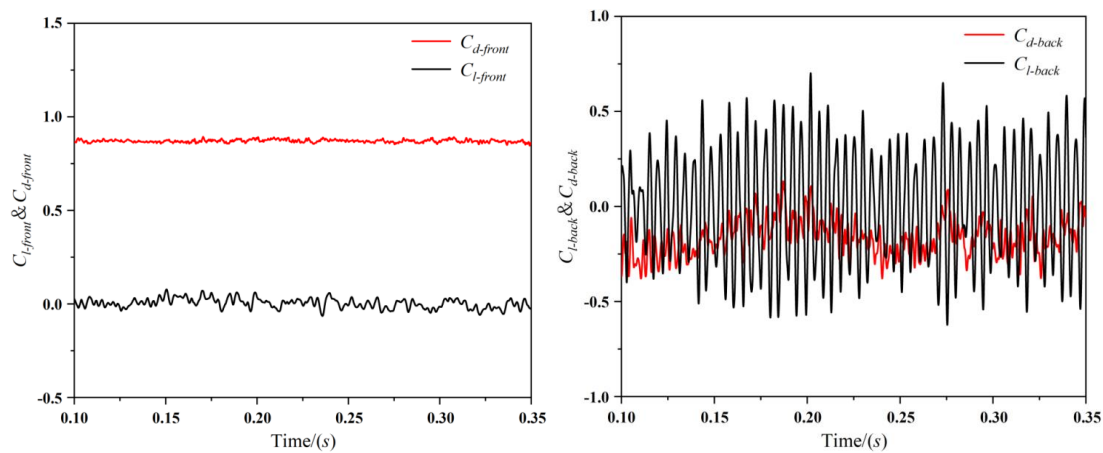

Fig 5-1 Lift and drag coefficient of front cylinder(left) and back cylinder (right)

Comparison of the lift and drag coefficients of the front and back cylinders shows that there are considerable fluctuations in the back cylinder compared to the front cylinder, with a large difference in the amplitude of the lift coefficients and a similarly large difference between the mean lift coefficient values. This is because the back cylinder completes the complete vortex shedding process and receives the turbulent energy from the front cylinder. In addition, the fluctuation amplitude of each vortex shedding cycle is also very different for Cd and Cl. The unsteady fluctuation of Cd is more significant in the back cylinder. The non-smooth fluctuations can be attributed to the strength of the vortex shedding from the front cylinder and the position of the impingement on the back cylinder over time. Similar phenomena have been found by Kitagawa and Ohta (2008)[37]; Sohankar (2014)[38]; and Gopalan and Jaiman (2015)[39] in their studies of tandem cylinders.

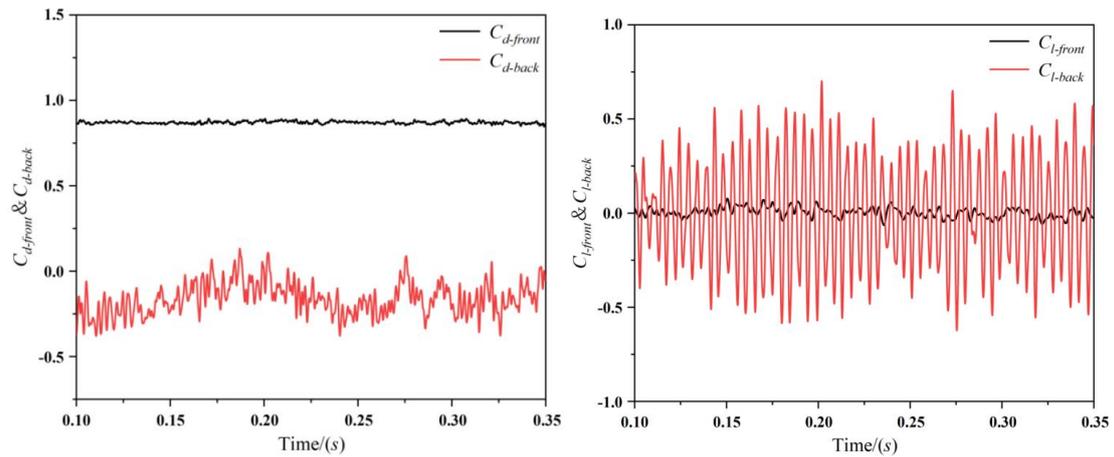

Fig 5-2 Comparison of drag(lift) and lift(right) coefficient of front and back cylinders

There is little difference between the elliptical cylindrical tandem winding lift and drag coefficient trend and that of the tandem cylindrical winding, but there is a large numerical difference.

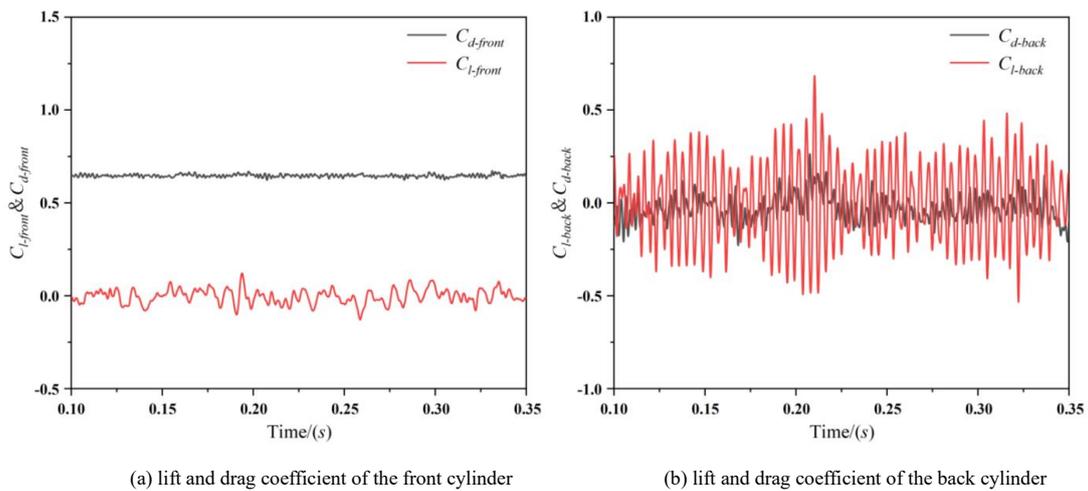

(a) lift and drag coefficient of the front cylinder   (b) lift and drag coefficient of the back cylinder

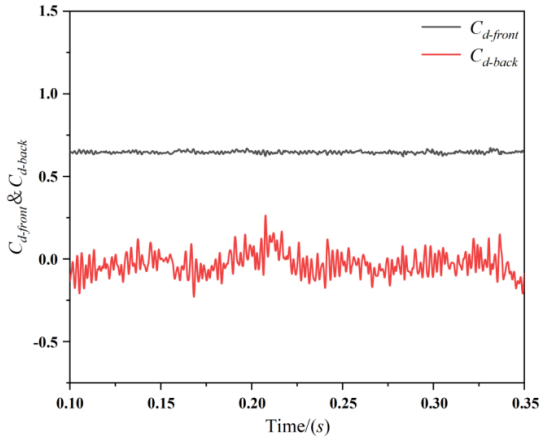 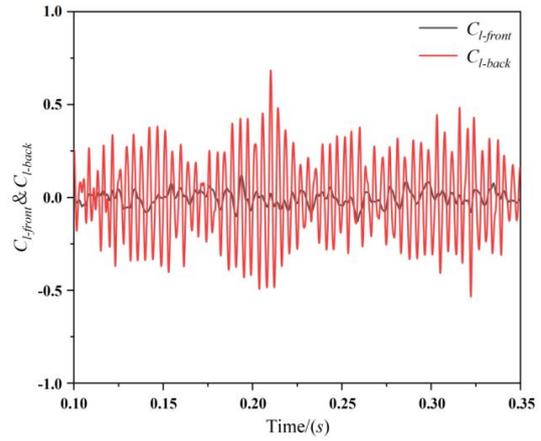

(c) drag coefficient of front and back elliptical cylinder    (d) lift coefficient of front and back elliptical cylinder

Fig 5-3 Comparison of lift and drag coefficient of front and back elliptical cylinder and cylinder

Comparing the lift drag coefficient of the front elliptical cylinder with that of the front cylinder, the drag coefficient of the elliptical cylinder is greatly reduced compared to that of the cylinder, which is about 0.225 or so. However, the amplitude of the lift coefficient is much larger for the elliptical cylinder than for the cylindrical one.

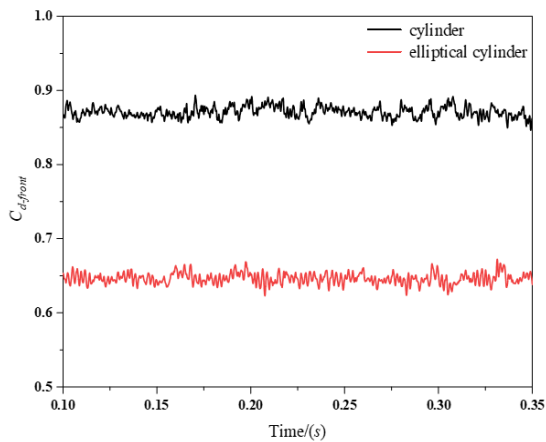 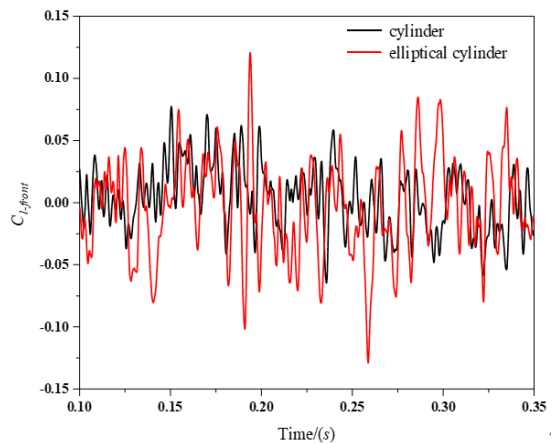

Fig 5-4 Comparison of drag(lift) and lift(right) coefficients of front cylinder and elliptical cylinder

Comparing the lift coefficients of the back cylinder with the back elliptical cylinder, the drag coefficients of the back elliptical cylinder increase in comparison to the cylinders, but the mean value is still less than 0. As for the amplitude of the lift coefficients, the magnitude of lift coefficients of the back elliptical cylinder decreases compared to that of the cylinders as a whole and fluctuates even less.

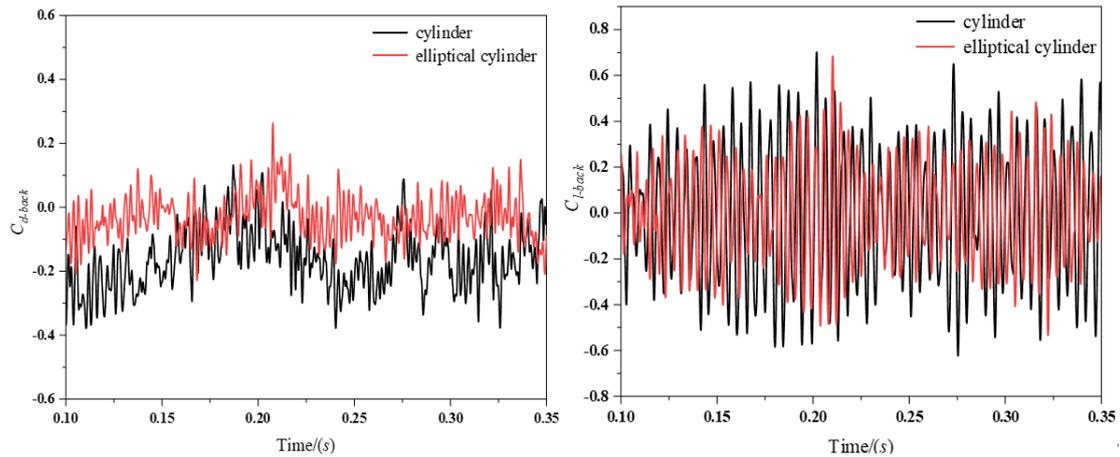

Fig 5-5 Comparison of drag(lift) and lift(right) coefficients for back cylinder (front elliptical/cylindrical respectively)

Fig 5-6 shows the comparison of the St number of the front cylinder and elliptical cylinder and the St number of the back cylinder (front of elliptical/cylindrical, respectively). Since the front did not produce a complete vortex shedding process, the St number is almost negligible, while for the back cylinder, the two produce a large gap, for the elliptical cylinder after the cylinder its St number is 0.155, and after the cylindrical St number is much higher with 208.8.

  Due to the fact that the aerodynamic analysis of a tandem double column is more complex and unpredictable compared to a single column. Therefore, in the force analysis section of the tandem double column, we will compare the front column of the tandem cylinder with the front column of the tandem elliptic column, and the back column of the tandem cylinder with the back column of the tandem elliptic column, respectively. And because in the analysis of single column in Chapter 4, we observed that the force component fluctuation is more stable at Re=3900. Therefore, in the comparison of the tandem columns, we compare and analyze them all at Re=3900. In the figure below (left), because of the difference in the shapes of the cylindrical and elliptical columns, the front column of the tandem elliptical column is subjected to a much smaller Cd force component than that of the tandem cylindrical column. But the force component of the front column of the tandem cylinder is suppressed more significantly. In the figure below (right), the fluctuation range of the tandem elliptic column's rear column is limited to a much smaller region, but the value of this difference is not large. This is because the shape of the posterior column is the same for both, both being cylindrical. The reason for the difference is that the shape of the front column is different, and there is a slight difference in the wake formed after the flow field passes through the front column.

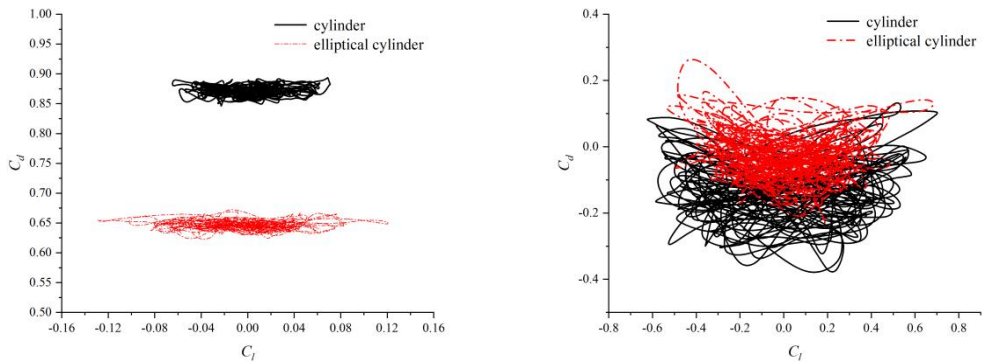

Comparison of the phase-space plot of the force coefficient components of front cylinder and elliptical cylinder(lift)

Comparison of the phase-space plot of the force coefficient components of back cylinder(front elliptical/cylindrical respectively)(right)

Observation of the following figure reveals that there is a significant difference between the shear stress cloud plots of the front column of the tandem column and the single column cloud plots. This is due to the inability of the front column to accomplish complete vortex shedding due to the influence of the rear column. The front column of the double tandem column forms a less obvious vortex structure but the elliptical column is relatively chaotic. Also, the shear stresses in the tandem elliptical columns are smaller relative to those in the tandem cylinders.

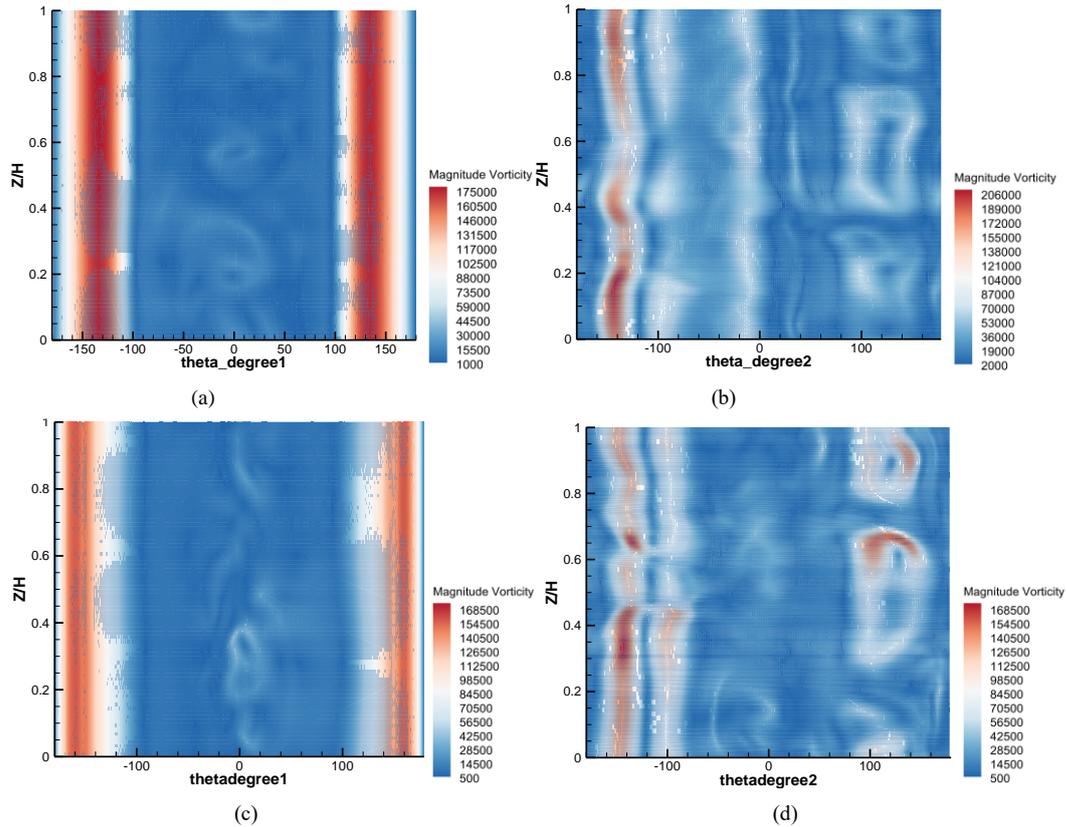

Fig The wall shear stress with $Re$=3900: (a) the front cylinder (b) the back cylinder

(c) the front elliptical cylinder (d) the front and back elliptical cylinder

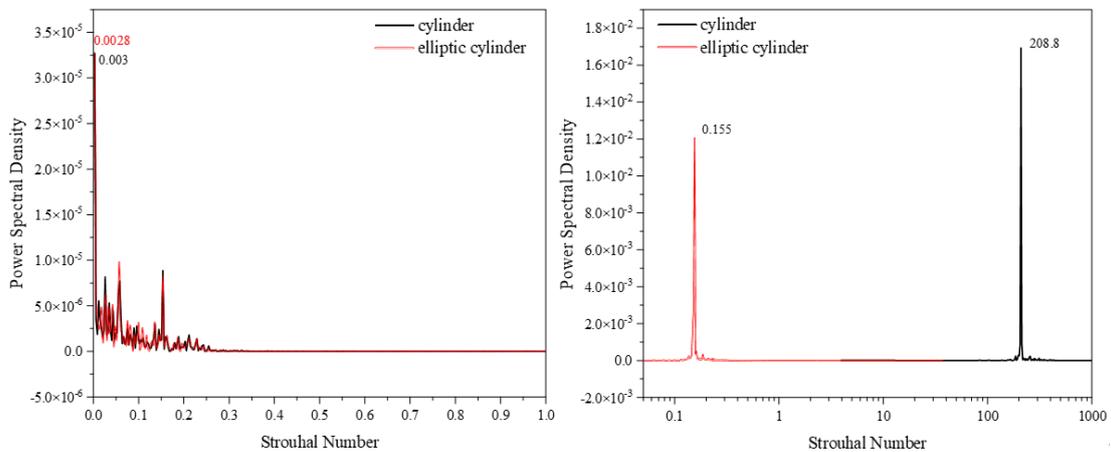

Fig 5-6 Comparison of St number of front cylinder and elliptical cylinder(lift)

Comparison of St number of back cylinder(front elliptical/cylindrical respectively)(right)

Fig 5-7 shows the comparison of the surface pressure coefficients of the front and back structures, respectively; Figure 5-7(left) shows the comparison of the surface pressure coefficients of the front cylinder and the elliptic cylinder, which shows that the pressure coefficients of the front surface of the elliptic cylinder change faster, with a larger gradient and a faster formation of the negative pressure, and the pressure coefficients of the surface of the elliptic cylinder are smaller than those of the cylinder, and the elliptic cylinder reaches the maximum negative pressure at about 50°, while the cylinder reaches the maximum point at about 70°. The maximum negative pressure is reached at about 50° for the elliptic cylinder, and at about 70° for the cylindrical cylinder. Fig 5-7(right) shows the comparison of the average pressure coefficients between the back cylinder and the elliptic cylinder, which shows that the pressure coefficients of the elliptic cylinder are significantly lower than those of cylinder in most of the regions and the inflection point of the pressure coefficients occurs earlier, the maximum pressure coefficient of the front surface of elliptic cylinders is smaller than that of cylinders, the maximum pressure coefficient of the rear surface of elliptic cylinders does not differ much from that of cylinders. The maximum pressure coefficient of the rear surface of the elliptic cylinder is not much different from that of the cylinder. In addition, the pressure coefficients on the surface of the elliptic cylinders are the same as those of flow around the single-cylinder, and the trend of the pressure coefficients on the surface of the cylinders is the wave type which decreases first, then increases, then decreases again, and then increases and decreases again.

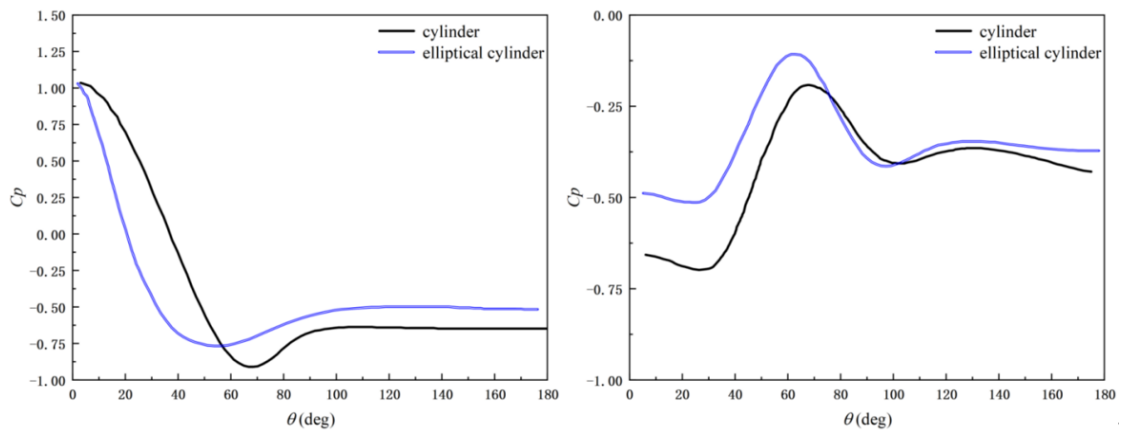

Fig 5-7 Comparison of average pressure coefficients on front elliptical and cylinder surfaces (left)

Comparison of average pressure coefficients on back cylinder surfaces (front elliptic/cylindrical, respectively) (right)

## 5.2 Flow Field Analysis

The velocity gradient of the elliptic cylinder is lower and the inflection point of zero velocity gradient occurs later for the elliptic cylinder compared to the cylindrical one, both in the flow field between the structures and in the flow field after the back cylinder. In addition, there are different points in the flow field before and after, i.e., in the flow field between the structures, the flow velocity in the elliptic-cylindrical case is smaller than that in the cylindrical case; for the flow field after the back cylindrical case, the flow velocity in the elliptic-cylindrical case is larger than that in the cylindrical case and then is smaller than that in the cylindrical case, which indicates that the elliptic-cylindrical case can reduce the velocity of the fluid effectively. It is worth noting that in the flow field between the elliptical cylinder and the cylinder, there is also a prominent point in the elliptical cylinder case, which reflects the different flow field characteristics of the elliptical cylinder.

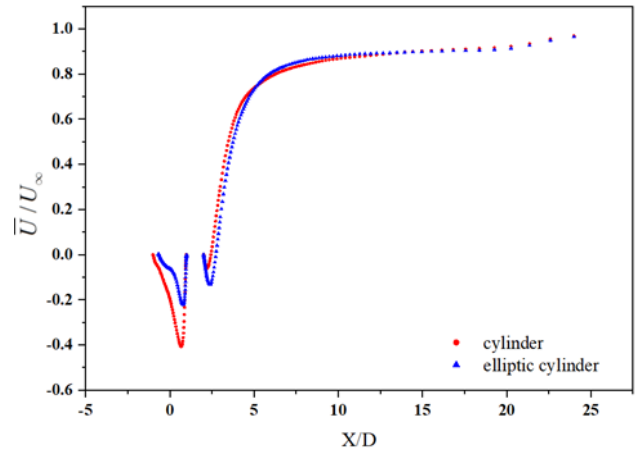

Fig 5-8 Comparison of time-averaged streamwise velocity Ux=U$_\infty$ (dimensionless quantity) along the centerline of the elliptical cylinder and the cylinder

Fig 5-9 and 5-10 show the vorticity and the end view vortical structures by the Q-criterion (Q = 1000), rendering at X-direction. In both cases, there is a clear periodic Carmen vortex structure behind the downstream cylinder. For the cylindrical case, the fluid flow velocity shedding from its walls is greater, and a comparison shows that the shear layer on the surface of the elliptical cylinder is more stable than the cylinder for both sets of Reynolds numbers and that the cylindrical case is unstable further downstream, transforming into small-scale vortex structures. The distribution of vortex structures in the flow field of the elliptic cylinder is more concentrated, and structures such as large-scale flow vortices and hairpin vortices are also significantly reduced. In addition, it can be seen from Figs. 5-7 that there are complex vortex structures between cylinders and cylinders and the fluid shedding from the cylinder is farther away and even wraps around the downstream cylinder, while for the elliptical cylinder case, the fluid shedding is much shorter in distance and completely reattaches to the front end of the downstream cylinder, which makes the vortex shedding from the cylinder downstream of the cylinder more intense, and the shear layer is more unstable, and the vortices are unstable downstream of the cylinder, which is transformed into small-scaled vortex structures.

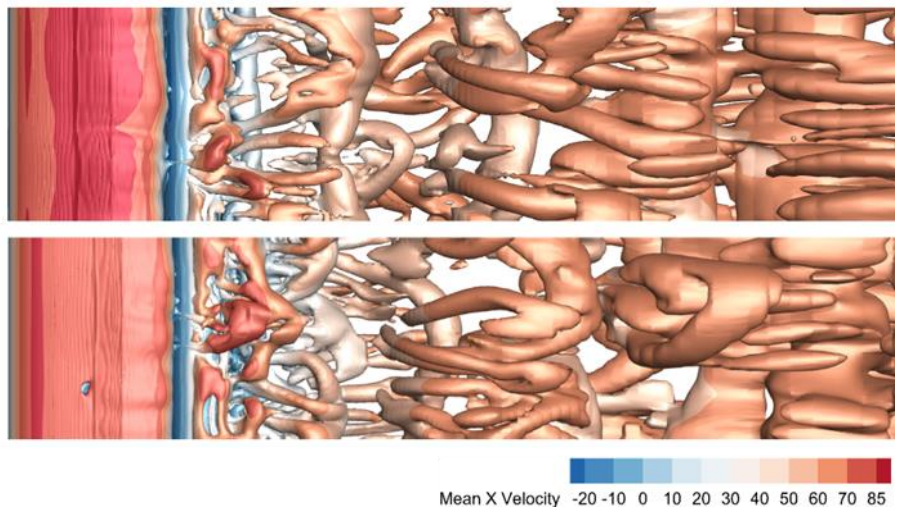

Fig5-9 The end view of the vortical structures by the Q-criterion(Q = 1000)

of the cylinder (former) and the elliptical cylinder(later)

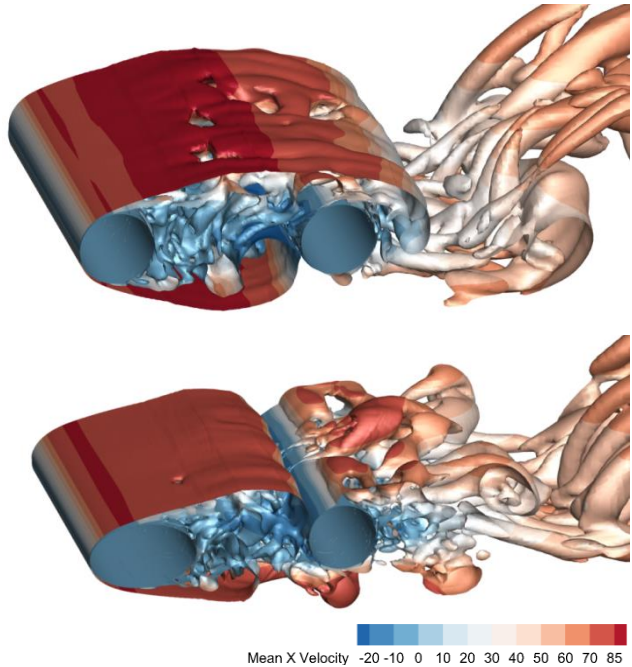

Mean X Velocity  -20 -10  0  10 20 30 40 50 60 70 85

Fig 5-10 The vortical structures by the Q-criterion(Q = 1000) of the cylinder (former) and the elliptical cylinder(later)

Fig 5-11 and 5-12 show the time-averaged pressure coefficient distributions on the surface of the cylinder/elliptic cylinder and its expansion for the two cases, and it can be seen that both surfaces have a large pressure gradient, and the pressure gradient of the elliptic cylinder is larger and the positive pressure zone is less, and the negative pressure zone is more; there is no pressure coefficient change on the rear surface of the cylindrical case, while the rear surface of the elliptic cylinder has different pressure coefficients distributed. For the back cylinder affected by the results of the front structure around the flow field, the distribution of pressure coefficients is quite different, with different sizes of pressure coefficients distributed at intervals on the rear surface of the back cylinder after the cylinder.

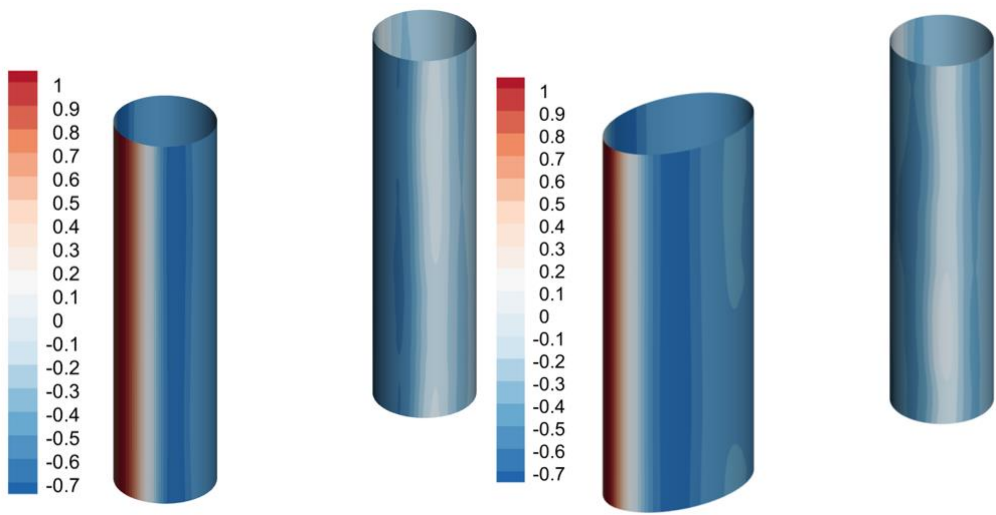

Fig 5-11 The 3D averaged pressure stress coefficient program of the cylinder (lift) and the elliptical cylinder(right)

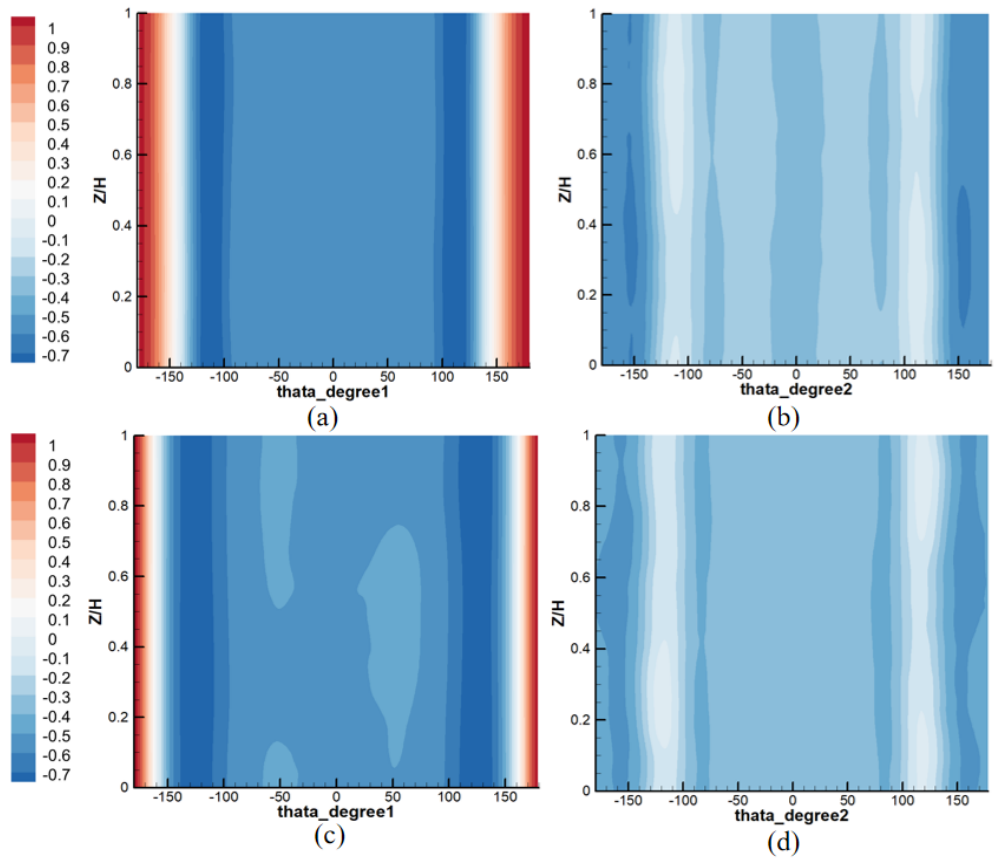

Fig 5-12 (a) front cylinder (b) back cylinder (c) front elliptical cylinder (d) back cylinder (front is elliptical)

the 2D averaged pressure stress coefficient program

Fig 5-13 and 5-14 show the two-dimensional distribution of dp/dt for the cylinder and elliptical cylinder, reflecting the rate of change of pressure, and it can be seen that the pressure gradient in the tail of the cylinder changes faster and are concentrated in the tail of the cylinder; for the elliptical cylinder, the distribution of the pressure change on the surface is more irregular and chaotic. For the rear cylinder in the conventional double-cylinder tandem configuration, vortex shedding on both sides of the cylinder results in a more pronounced variation in the rate of change of positive and negative pressure. In the case of the elliptical cylinder, the surface also exhibits a more chaotic and asymmetric distribution of the rate of change of pressure.

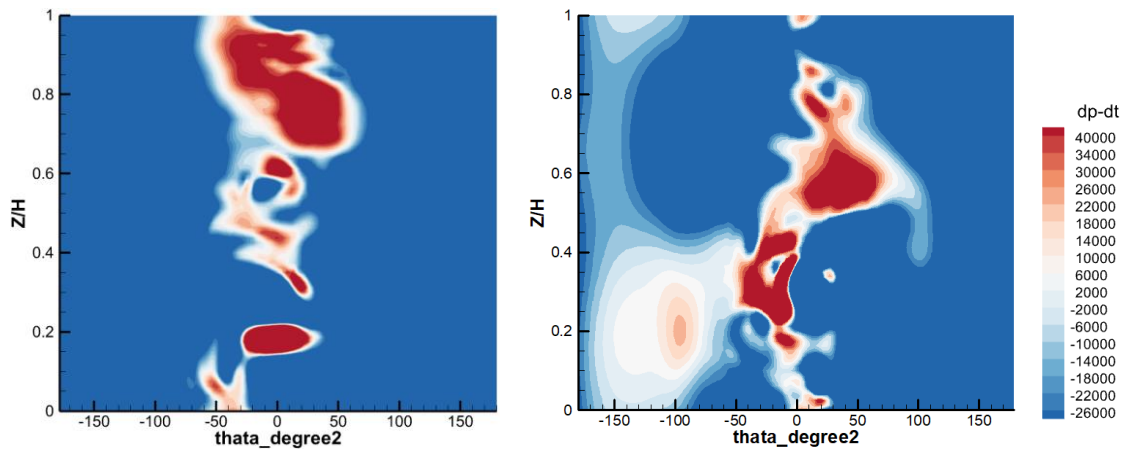

Fig 5-13 develop diagram of 2D surface of front cylinder(lift) and back cylinder(right)

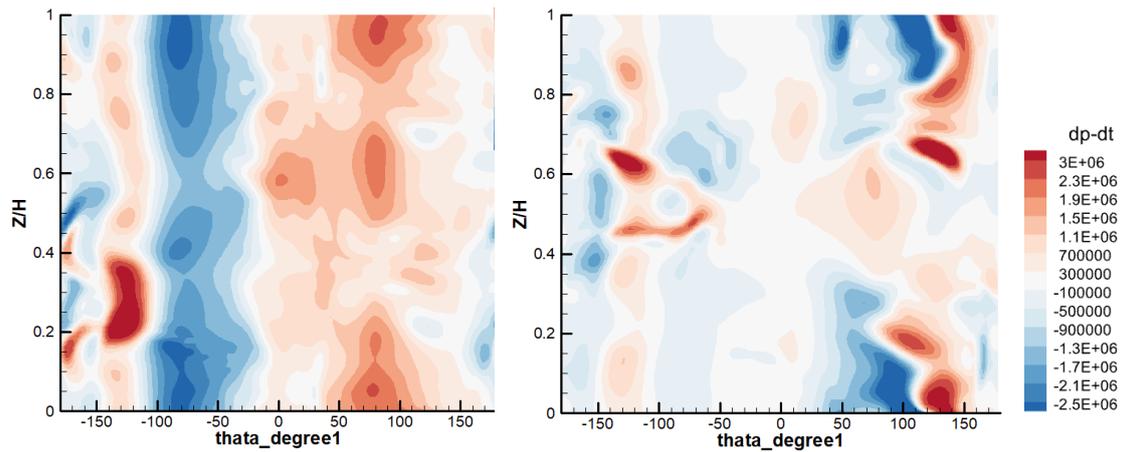

Fig 5-14 develop diagram of 2D surface of front elliptical cylinder(lift) and back elliptical cylinder(right)

Comparison of the surface-limiting streamlines of cylinder and elliptic cylinder tandem windings shows that the surface-limiting streamlines of elliptic cylinders are more regular than the spreading distributions of cylinders due to the controlling effect of elliptic cylinders. In addition, a completely different streamline distribution exists for the back cylinders influenced by the front structural body, where the elliptical post-cylindrical surface-limit streamlines distribution is more regular and sparser on the forward flow-meeting surface, but more chaotic on the rear surface than in the cylindrical tandem case, with a non-uniform distribution at the spreading height. Finally, the separation lines are the same in both cases and are in approximately the same locations.

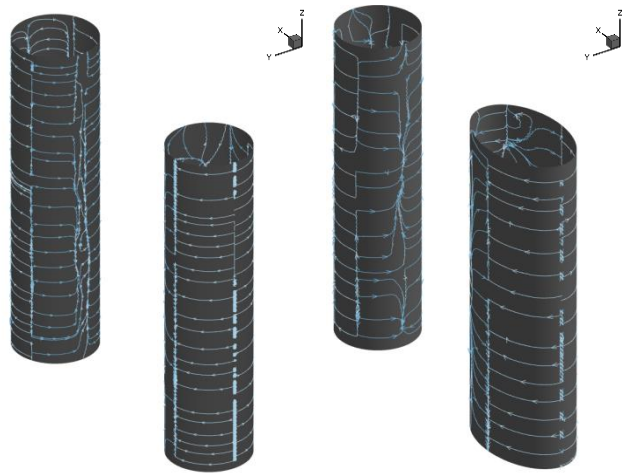

Fig 5-15 The front view of the surface limiting streamline distribution of the cylinder and the elliptical cylinder

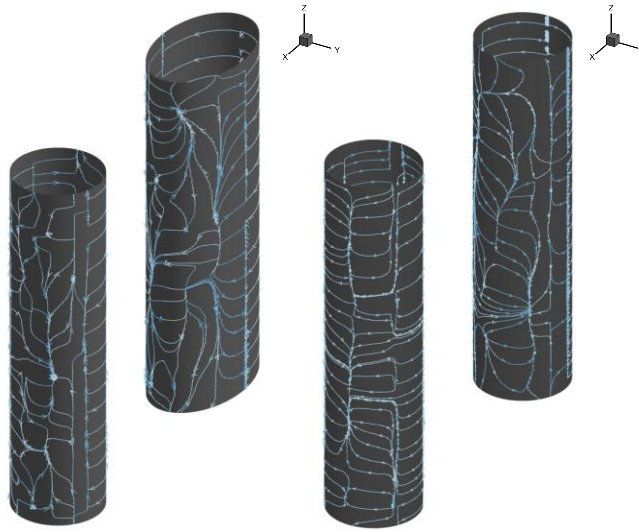

Fig 5-16 The back view of the surface limiting streamline distribution of the cylinder and the elliptical cylinder

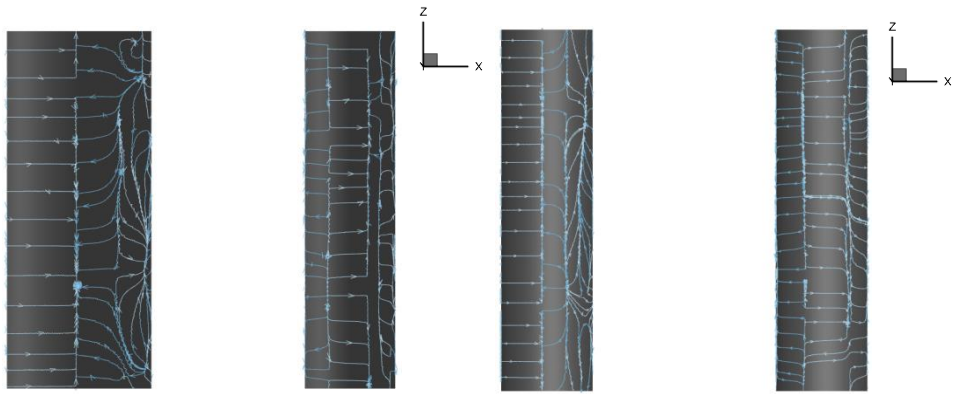

Fig 5-17 The end view of the surface limiting streamline distribution of the cylinder and the elliptical cylinder

Fig 5-18 and 5-19 show the dimensionless 2D vortex structure and end views for the two cases. The vortices are larger in the cylindrical case compared to the elliptical case and there are more vortex structures and larger vortices in the flow field between the structures. The cylindrical vortex shedding after the elliptical column is longer and fewer in number.

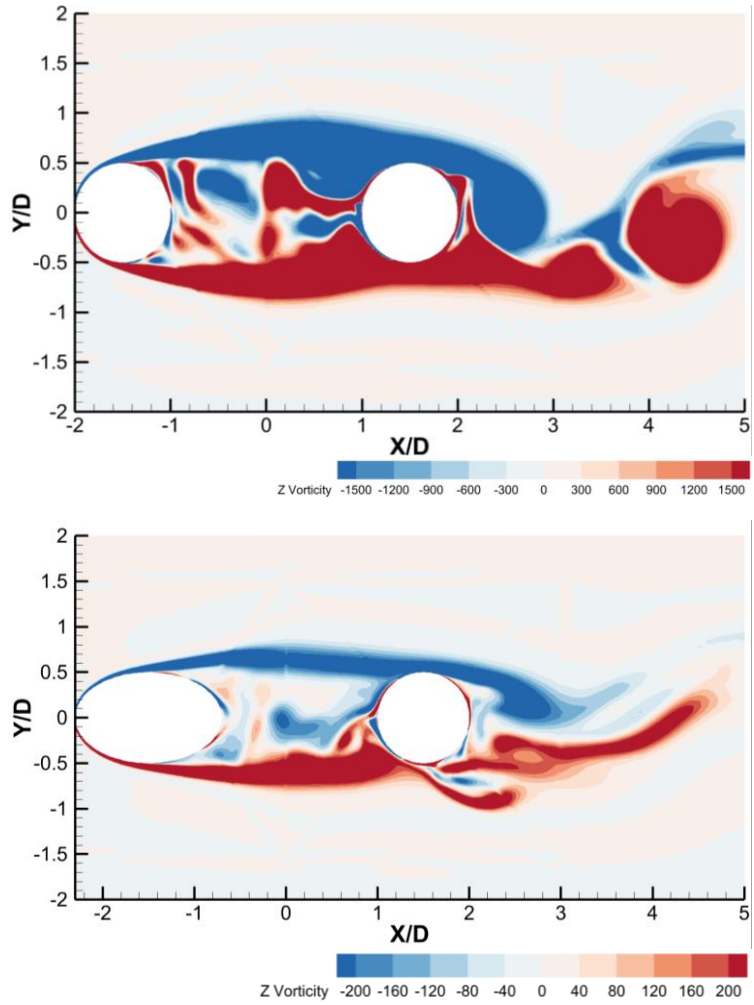

Fig 5-18 The 2D vortex structure of the cylinder(former) and the elliptical cylinder(later)

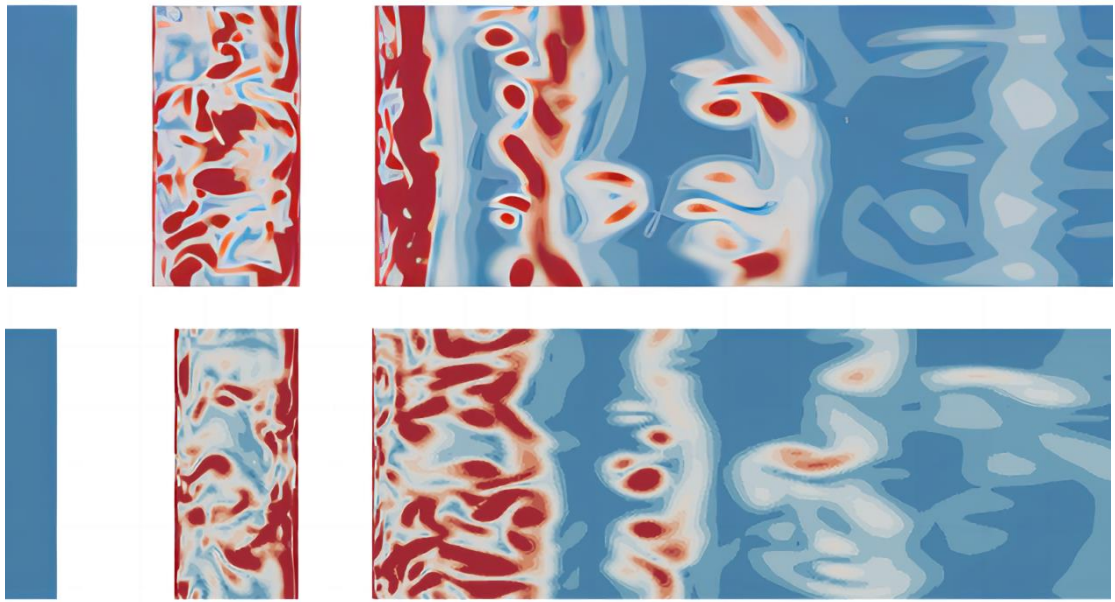

Fig 5-19 The end view of the 2D vortex structure of the cylinder(former) and the elliptical cylinder(later)

Fig 5-20(a) shows the dimensionless 2D x-direction time-averaged velocities for the two cases, and the high-speed region on both sides of the structure is larger and wider in the cylindrical case, and the front end of the back cylinder shows a higher negative velocity of -20m/s, and the flow field is more symmetric in the cylindrical case. Fig 5-20(b) shows the dimensionless 2D streamlines for the two cases, and it can also be seen that the streamline region on both sides of the structure is larger and wider in the cylindrical case, and the flow field is more symmetric in the cylindrical case. It is worth noting that the range of cylindrical streamlines in the elliptical-cylindrical case has a larger and wider range and is more widely derived along the flow direction.

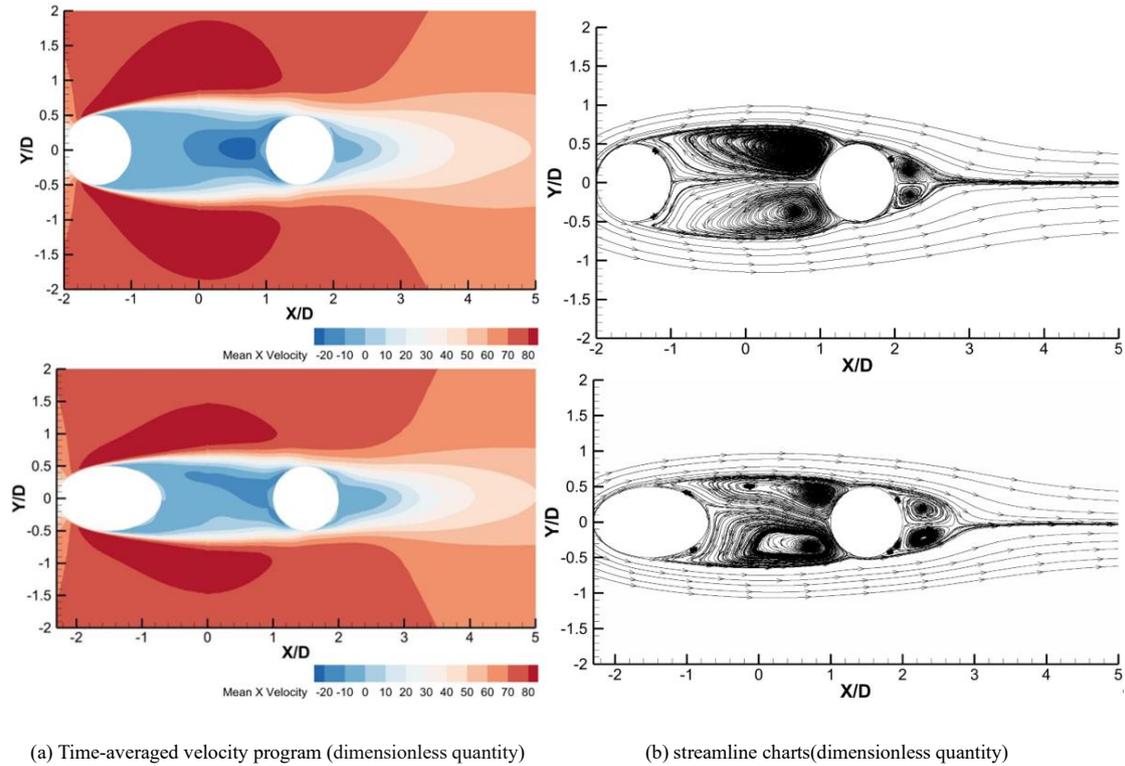

(a) Time-averaged velocity program (dimensionless quantity)　　　(b) streamline charts(dimensionless quantity)

Fig 5-20 Time averaged velocity and streamlines

## 5.3 Analysis of Far-field Radiated Noise

Figs 5-21, 5-22, and 5-23 show the frequency dependence of the series cylindrical and series elliptical noise SPLs at each observation point from 24 observation points uniformly distributed around the area. Each observation point elliptical column noise SPL is overall smaller than the cylinder and can be found in addition to the flow direction of 0 ° and 180 °, elliptical cylinder series noise main frequency is slightly larger than the cylinder, but the maximum SPL is smaller than the cylinder.

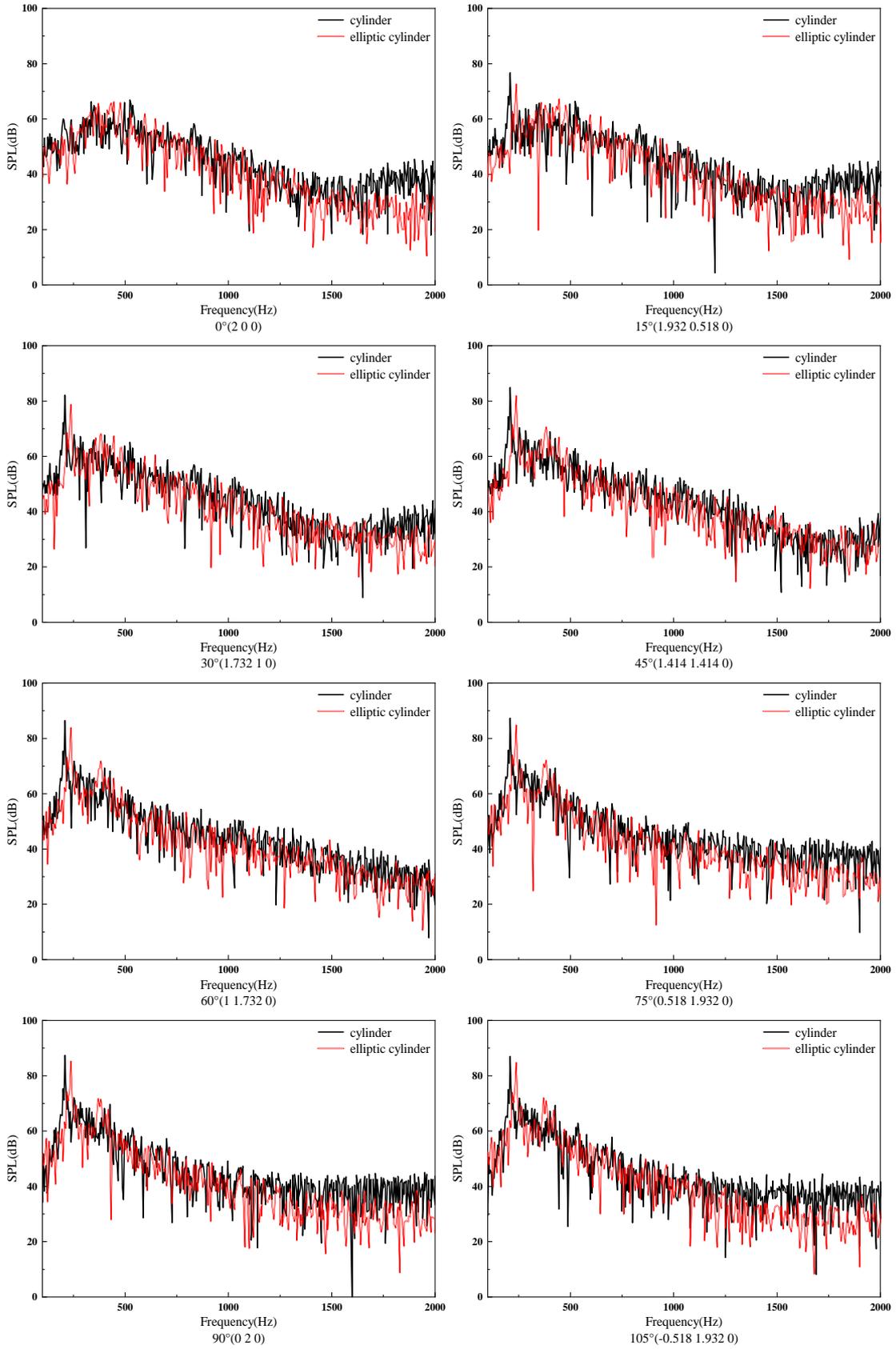

Fig 5-21 Comparisons of acoustic pressure spectra with the frequency of the cylinder and the elliptical cylinder

the observation points 1-8

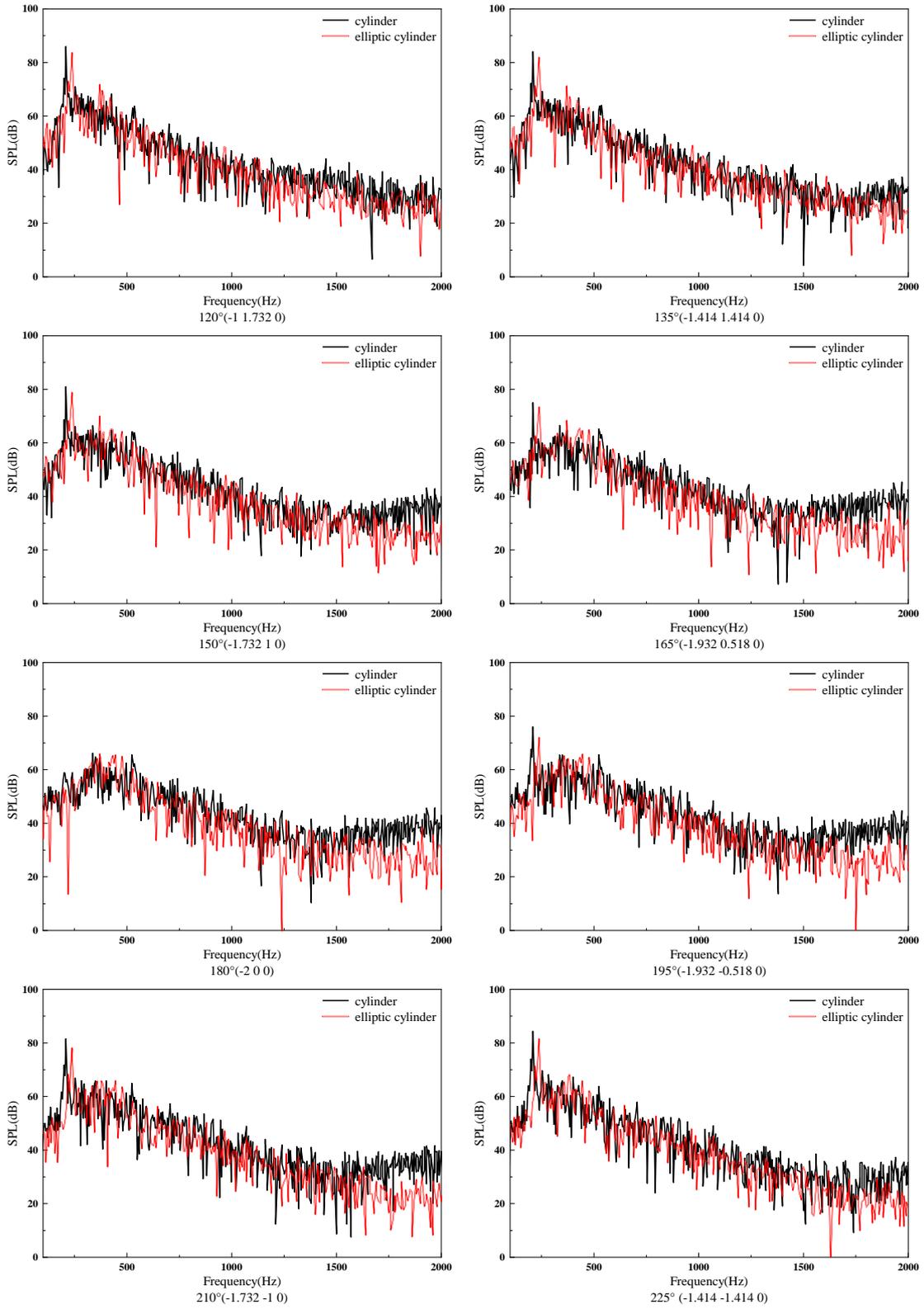

Fig 5-22 Comparisons of acoustic pressure spectra with the frequency of the cylinder and the elliptical cylinder observation points 9-16

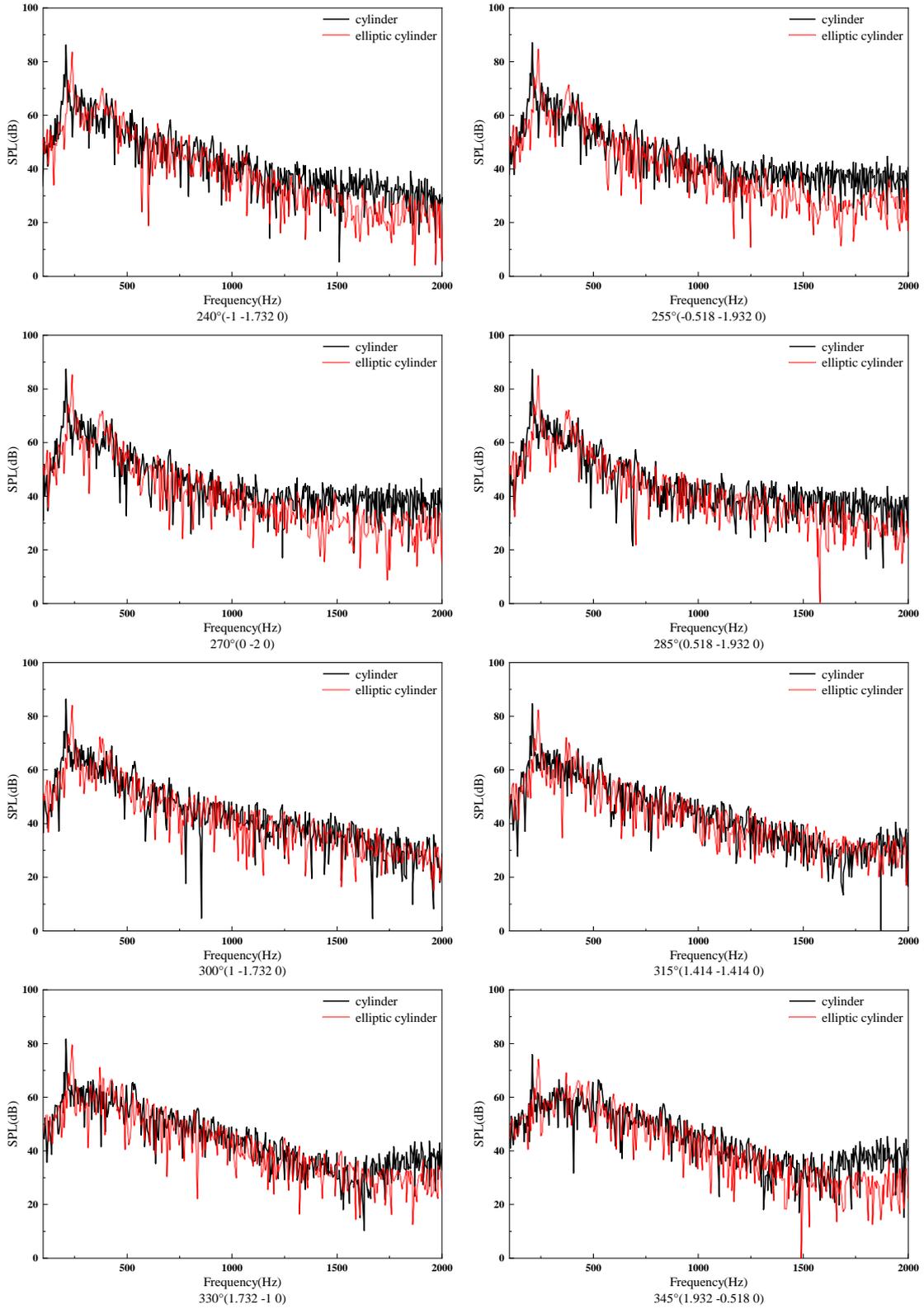

Fig 5-23 Comparisons of acoustic pressure spectra with the frequency of the cylinder and the elliptical cylinder

the observation points 17-24

Same to the single cylinder analysis, based on 24 uniform distribution observation points along the perimeter of the circle formed by the two structures as the origin and 40D as the radius, the OASPL of the total sound pressure level of each observer is obtained through the calculation and the sound directivity distribution map shown in Fig 5-24 is realized. The shape of the noise directivity is the same in both cases, and both have dipole modes. It is worth mentioning that the elliptic cylinder can significantly reduce the noise radiation in all directions. To explore the noise reduction effect in each direction, Fig 5-25 shows the curve of the noise reduction amount with the change of the angle. It can be seen that the noise reduction amount shows the tendency of increasing and then decreasing, and the noise reduction amount is largest in the 180 °, i.e., at the point of the flow on the stream, which is up to 4.5 dB.

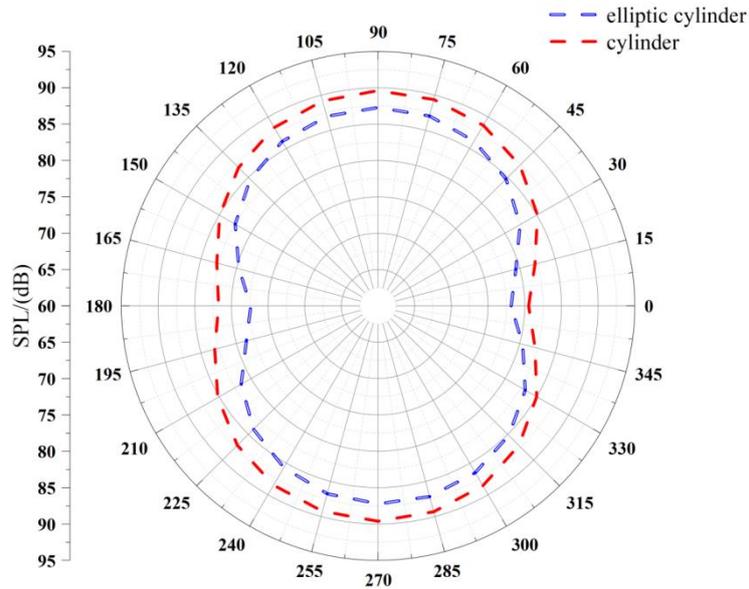

Fig 5-24 Comparison of the directivity of the noise of the cylinder and the elliptical cylinder

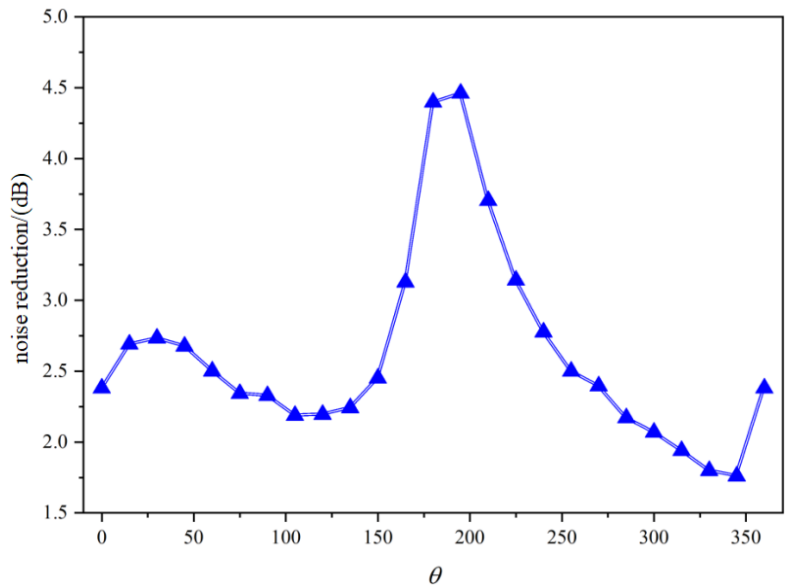

Fig 5-25 Variation of noise reduction along the surface of the cylinder

# 6. Conclusion

The influence of the Reynolds number and configuration on the flow field and aerodynamic noise is studied by combining IDDES and FW-H, and the conclusion is obtained as follows:

In the high Reynolds number case, the lift coefficient fluctuates more and the overall lift coefficient is larger, the resistance coefficient is larger, and the St number is larger so that the vortex falls off faster. The separation position of the cylinder surface moves downward, the surface shear layer is more unstable, and the core region of the turbulence is located further upstream of the wake region. The distribution of vortex structure in the low Reynolds number flow field of cylinder is more concentrated, and the structure of large-scale flow vortex and hairpin vortex is also significantly reduced, and the vortex strength of wake is significantly weakened. Noise radiation at a higher Reynolds number increases in all directions, with the most pronounced increase in noise at 0° and 180° and the least in directions perpendicular to the flow direction.

The elliptic column can significantly reduce the noise radiation in all directions, and the noise reduction shows a trend of first increasing and then decreasing along with the surface of the cylinder. At 180°, the current point, the noise reduction is the largest, which can reach 4. 5dB. The non-stationary fluctuations of the drag coefficient are more pronounced on the downstream cylinder. Compared with the upstream cylinder, the lift coefficient amplitude is much larger than the cylinder. The elliptical column can effectively reduce the velocity of the flow through the fluid. In the cylindrical case, the vorticity is larger than in the elliptical case, and the vortex structure is more finely broken. The noise sound pressure level of the elliptical cylinder at each observation point is smaller than the cylinder. In both cases, the noise has good directivity, both in dipole mode.